\documentclass[conference]{IEEEtran}
\IEEEoverridecommandlockouts
\usepackage{cite}
\usepackage{amsmath,amssymb,amsfonts}
\usepackage{algorithmic}
\usepackage{graphicx}
\usepackage{textcomp}
\usepackage{xcolor}
\usepackage{comment}
\usepackage{cleveref}
\usepackage{verbatimbox}

\usepackage{caption}
\usepackage{latexsym}
\usepackage{bbding}
\usepackage{bm}
\usepackage{array}
\usepackage[ruled,vlined,linesnumbered]{algorithm2e}
\usepackage{subfigure}
\usepackage{multirow}
\usepackage{makecell}
\usepackage{amsmath}

\usepackage{enumitem}
\usepackage{diagbox}
\usepackage[font=footnotesize]{caption}
\renewcommand{\baselinestretch}{0.972}
\usepackage{setspace}
\usepackage{url}
\makeatletter
  \newcommand\figcaption{\def\@captype{figure}\caption}
  \newcommand\tabcaption{\def\@captype{table}\caption}
\makeatother

\usepackage{xcolor}

\def\BibTeX{{\rm B\kern-.05em{\sc i\kern-.025em b}\kern-.08em
    T\kern-.1667em\lower.7ex\hbox{E}\kern-.125emX}}
\begin{document}

\title{Semantic Guided and Response Times Bounded Top-k Similarity Search over Knowledge Graphs}

\author{
\IEEEauthorblockN{Yuxiang Wang\textsuperscript{1,2}, Arijit Khan\textsuperscript{2}, Tianxing Wu\textsuperscript{2},
Jiahui Jin\textsuperscript{3}, Haijiang Yan\textsuperscript{1}}
\IEEEauthorblockA{\textsuperscript{1} \textit{Hangzhou Dianzi University, China}
\textsuperscript{2} \textit{Nanyang Technological University, Singapore}
\textsuperscript{3} \textit{Southeast University, China}}
\{lsswyx,yanhj\}@hdu.edu.cn, \{arijit.khan,wutianxing\}@ntu.edu.sg, jjin@seu.edu.cn
}

\maketitle

\newtheorem{myExample}{Example}
\newtheorem{myDef}{Definition}
\newtheorem{myTheorem}{Theorem}
\newtheorem{myLemma}{Lemma}
\renewcommand{\IEEEQED}{\IEEEQEDopen}
\def\IEEEproofindentspace{2\parindent}
\renewcommand{\IEEEproofindentspace}{10pt}
\newcommand{\tabincell}[2]{\begin{tabular}{@{}#1@{}}#2\end{tabular}}

\begin{abstract}
Recently, graph query is widely adopted for querying knowledge graphs. Given a query graph $G_Q$, the graph query finds subgraphs in a knowledge graph $G$ that exactly or approximately match $G_Q$. We face two challenges on graph query: (1) the structural gap between $G_Q$ and the predefined schema in $G$ causes mismatch with query graph, (2) users cannot view the answers until the graph query terminates, leading to a longer system response time (SRT). In this paper, we propose a semantic-guided and response-time-bounded graph query to return the top-k answers effectively and efficiently. We leverage a knowledge graph embedding model to build the semantic graph $SG_Q$, and we define the path semantic similarity ($pss$) over $SG_Q$ as the metric to evaluate the answer's quality. Then, we propose an A* semantic search on $SG_Q$ to find the top-k answers with the greatest $pss$ via a heuristic $pss$ estimation. Furthermore, we make an approximate optimization on A* semantic search to allow users to trade off the effectiveness for SRT within a user-specific time bound. Extensive experiments over real datasets confirm the effectiveness and efficiency of our solution.
\end{abstract}

\section{Introduction}
Knowledge graphs (such as DBpedia \cite{Mendes2012}, Yago \cite{Hoffart2013}, and Freebase \cite{Bollacker2007}) have been constructed in recent years, managing large-scale and real-world facts as a graph \cite{Huang2019}. In such graphs, each node represents an entity with attributes, and each edge denotes a relationship between two entities. Querying knowledge graphs is essential for a wide range of applications, e.g., question answering and semantic search \cite{Guha2003}. For example, consider that a user wants to find
\textit{all cars produced in Germany}. One can come up with a reasonable graph representation of this query as a query graph $G_Q$, and identify the exact or approximate matches of $G_Q$ in a knowledge graph $G$ using graph query models  \cite{Khan2013,Zou2014,Yang2014,Yang2016,Jin2015}. Correct answers can be returned, such as $\langle$\textit{BMW\_320}, \textit{assembly}, \textit{Germany}$\rangle$. Graph query also acts as a fundamental component for other query forms, such as keyword and natural language query \cite{Yang2016}. We can reduce these query forms to a graph query by translating input text to a query graph \cite{Zheng2015a,Han2017}.

To retrieve the information of interest from a knowledge graph $G$, users are often required to have full knowledge of the vocabulary used in $G$ \cite{Shekarpour2017}, as well as the underlying schemas defined in $G$, which is difficult for ordinary users (even professional users) \cite{Zheng2016}. Otherwise, the user-built query graph is likely to be structurally different from the predefined schemas, thus fails to return correct answers due to the mismatch with the query graph. Consider the following motivating example.

\begin{figure}
\setlength{\abovecaptionskip}{0cm}
\setlength{\belowcaptionskip}{-0.7cm}
\centerline{\includegraphics[scale=0.5]{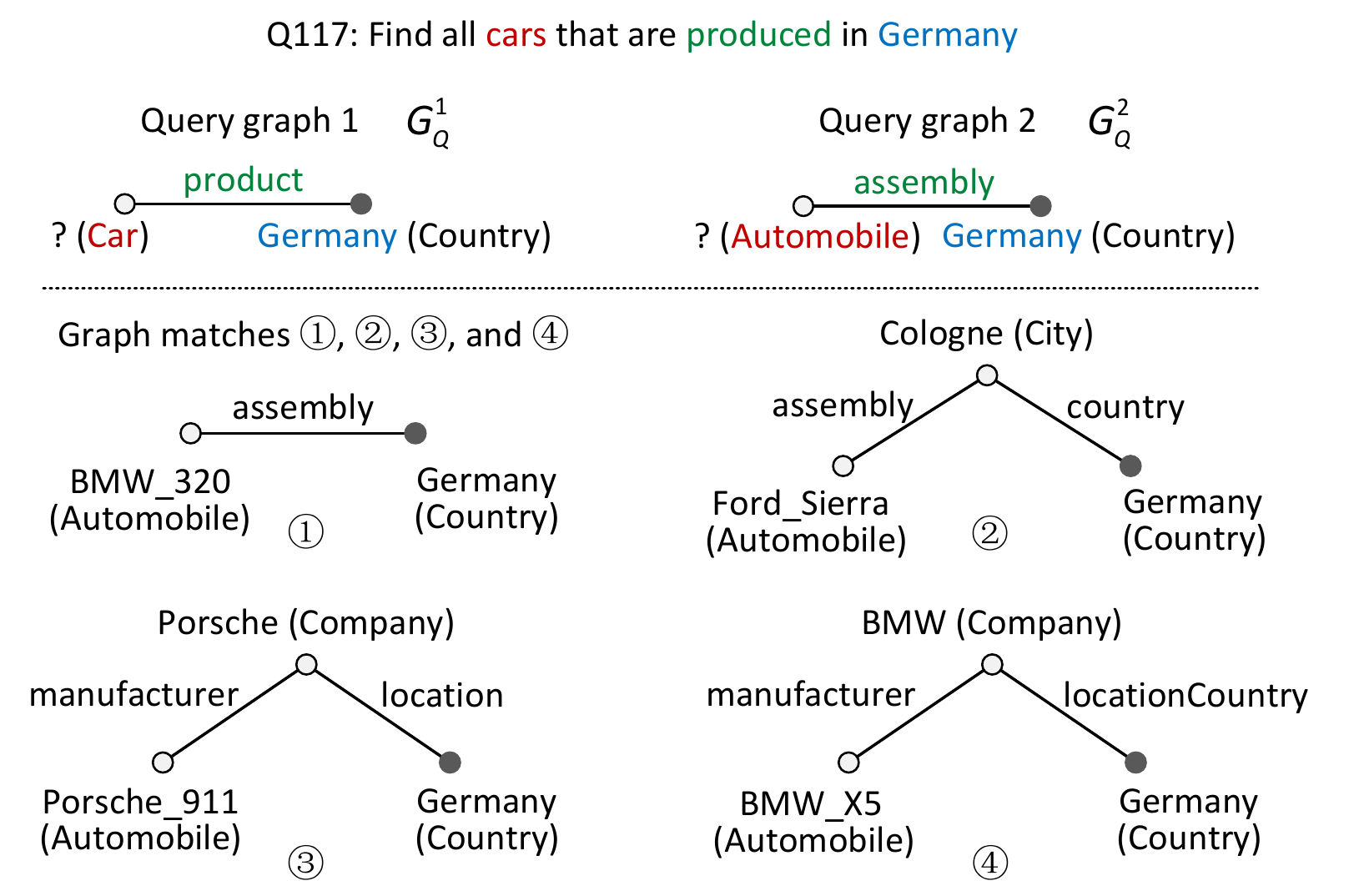}}
\caption{An example of structural mismatch with query graphs: different users may employ different query graphs (top) to find all cars made in Germany. Four correct graph matches in DBpedia are provided (bottom). Only $G_Q^2$ can retrieve partial correct answers having the same schemas as \textcircled{1}, because the 1-hop edge \textit{assembly} cannot match any $n$-hop ($n>1$) paths.}
\label{fig:example2}
\end{figure}

\begin{myExample}
Consider the query: \textit{Find all cars that are produced in Germany} (Q117 from QALD-4 benchmark \cite{qald}). Figure \ref{fig:example2} provides four correct graph matches with different schemas in DBpedia. Each one is represented as an $n$-hop path. An ordinary user may build a query graph $G^1_Q$ based on the phrases from the natural language question \cite{Han2017}. While a professional user may build $G^2_Q$ by using the controlled vocabulary (e.g., \textit{Automobile} is used to represent vehicles in DBpedia) and a schema she already knew. However, both query graphs suffer from the structural mismatch problem.
\end{myExample}

\noindent\underline{\textbf{Mismatch in query nodes}}.
In $G_Q^1$, a query node with type \textit{Car} represents the phrase ``cars". However, no entity in DBpedia has the same, or even a textually similar type for \textit{Car}, because it is not a term belonging to the controlled vocabulary of DBpedia. Hence, $G^1_Q$ fails to find correct answers.

\noindent\underline{\textbf{Mismatch in query edges}}.
For $G^2_Q$, the user can retrieve 234 answers that have the same schema as graph match 1. However, more than 200 correct answers are ignored, because a 1-hop edge in $G^2_Q$ cannot be mapped to the semantically similar $n$-hop ($n>1$) paths (edge-to-path mapping).


If a user has full knowledge about DBpedia, then she can build various query graphs that cover all possible schemas, to obtain all cars of interest. Generally, it is a strong assumption. As an alternative, we aim to provide a graph query system that will be able to support different query graphs without forcing users to use very controlled vocabulary or be knowledgeable about the dataset. This motivates us to fill the structural gap in graph matching by considering the semantics of query graphs.

\begin{figure*}
\setlength{\abovecaptionskip}{0.1cm}
\setlength{\belowcaptionskip}{-0.6cm}
\centerline{\includegraphics[scale=0.48]{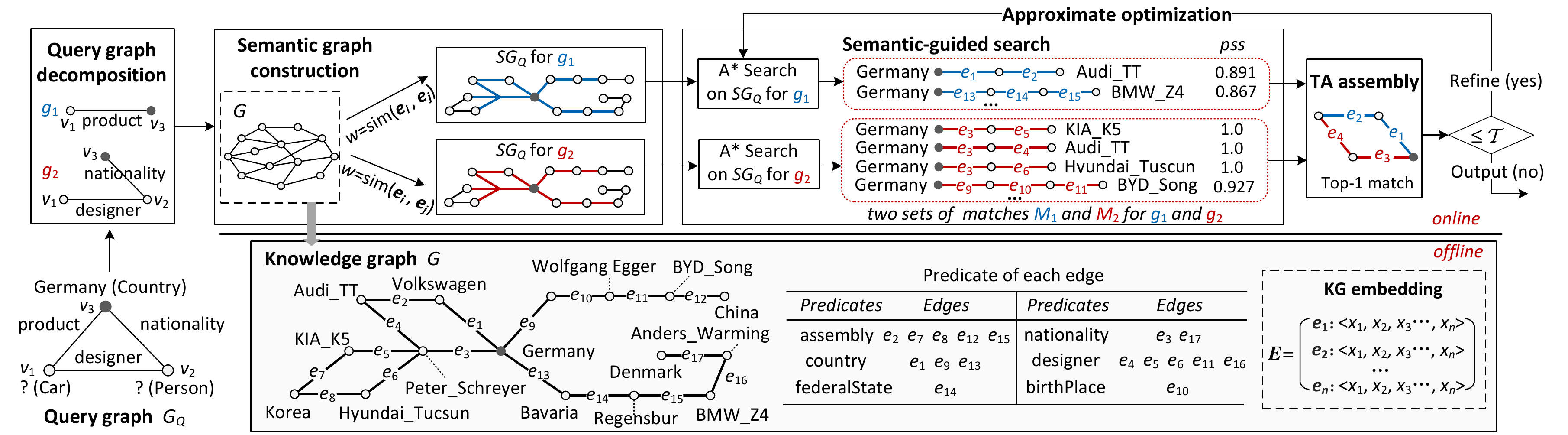}}
\caption{A running example of our approach, including offline KG embedding (bottom right) and online query processing. Four components of the online part are: (1) Query graph decomposition, (2) Semantic graph construction, (3) Semantic-guided search, and (4) Threshold Algorithm-based assembly. An approximate optimization is applied for response-time-bounded query. All predicates in the example knowledge graph are provided in a table (bottom middle).}
\label{fig:fullexample}
\end{figure*}

Another crucial problem involves improving the system response time (SRT) for a graph query. SRT is the amount of time that a user waits before viewing results \cite{Bhowmick2013}. A shorter SRT usually indicates a better user experience. To the best of our knowledge, no current state-of-the-art work supports response-time-bounded graph query.  This motivates us to present an interactive paradigm that allows the user to trade off accuracy for SRT within a user-specific time bound $\mathcal{T}$. As more time is given, better answers can be returned.

In this paper, we blend semantic-guided and response-time-bounded characteristics in one system to support the top-k graph query over a knowledge graph effectively and efficiently.


\vspace{-0.2cm}
\subsection{Our Approach}
\label{approach}
\vspace{-0.05cm}
Many efforts have been made for structural mismatch problem \cite{Yang2014,Khan2013,Fan2010a,Zheng2016,Jin2015,Han2017}. Among these methods, SLQ \cite{Yang2014} is the best one for the query node mismatch via a node transformation library, but it cannot support the edge-to-path mapping. S4 \cite{Zheng2016} is the most similar work to our paper. It mines the $n$-hop schemas in advance through \textit{string edit distance} and \textit{frequent paths}, by providing semantic instances as the prior knowledge (e.g., given by PATTY \cite{Nakashole2012}). It has two limitations, that are: (1) the string edit distance cannot well represent the real semantics of mined schemas, (2) the accuracy of S4 is sensitive to the quality of prior knowledge.

Unlike S4, we present a semantic-guided search to find the semantically similar paths to query edges without prior knowledge. To the best of our knowledge, we are the first to support semantically edge-to-path mapping in graph query. Moreover, combining our method with SLQ allows us to handle mismatches in query nodes. Figure \ref{fig:fullexample} shows the pipeline of our approach, which has an offline and an online phase.

\vspace{0.1cm}
\noindent\underline{\textbf{Offline phase}}. Given a knowledge graph $G$, we leverage a knowledge graph embedding model to represent the predicates of $G$ in a vector space $\bm{E}$ (\textbf{Section \ref{embedding}}). Hence, the semantic similarities of predicates can be easily obtained through vector calculation, and this makes it possible to achieve the semantic similarities of a path in $G$ to a query edge in $G_Q$. To be precise, this is essential for semantically edge-to-path mapping.

\noindent\underline{\textbf{Online phase}}. In Figure \ref{fig:fullexample}, we take a query graph $G_Q$ (that can be of different shapes \footnote{According to \cite{Bonifati2017}, chain, star, tree, cycle, and flower are the commonly used graph shapes.}) as an input. In this paper, we adopt a decomposition-assembly framework for $G_Q$, which consists of four basic components.

\vspace{0.1cm}
\noindent(1) \textit{Query graph decomposition}. We first decompose $G_Q$ into a set of sub-query graphs $\{g_1...g_n\}$ by a dynamic programming algorithm, subject to minimizing the possible query cost. Since this part is not our main focus in this paper, we briefly introduce it in \textbf{Section \ref{pre_overview}}, and emphasize more on the querying of the sub-query graphs and assembling their results.

\vspace{0.1cm}
\noindent(2) \textit{Semantic graph construction}. To support the semantically edge-to-path mapping, we then construct a semantic graph $SG_Q$ for each sub-query graph $g_i$ by preserving the predicate semantic similarities on the edges of $G$ (\textbf{Section \ref{weight}}). In Figure \ref{fig:fullexample}, we show example semantic graphs for $g_1$ and $g_2$. For instance, each blue edge has a high semantic similarity to the query edge \textit{product}, e.g., $e_2$ (\textit{assembly}) has 0.98 semantic similarity to \textit{product}. By utilizing these similarities in $SG_Q$, we define the path semantic similarity ($pss$) to measure how semantically similar a path in $G$ is to a sub-query graph $g_i$ (\textbf{Section \ref{pss}}).

\vspace{0.1cm}
\noindent(3) \textit{Semantic-guided search}. We next present an A* semantic search to find the top-k semantically similar matches for each sub-query graph $g_i$ from the semantic graph $SG_Q$ based on the path semantic similarity ($pss$). To improve the efficiency, we propose a well-designed heuristic estimation function of $pss$ that prunes the search space (\textbf{Section \ref{estimate}}). We prove the effectiveness guarantee of our A* semantic search in \textbf{Section \ref{Asearch_overview}}, that is, the matches with the greatest $pss$ must be found.

\vspace{0.1cm}
\noindent(4) \textit{Threshold Algorithm (TA)-based assembly}. Finally, we assemble the matches of all sub-query graphs based on the threshold algorithm (TA) \cite{Fagin2001}, in order to form the final answers for $G_Q$ in \textbf{Section \ref{final}}.

\vspace{0.1cm}
Moreover, we present an approximate optimization on our semantic-guided search to enable a trade-off between accuracy and the system response time (SRT) within a user-specified time bound $\mathcal{T}$ (\textbf{Section \ref{bound}}). As more time is given, more high-quality matches are refined incrementally. We prove that the globally optimal matches for $G_Q$ can be achieved theoretically when sufficient time is given.

\subsection{Contributions}
\label{challenges}
We summarize our main contributions as follows.
\begin{itemize}[leftmargin=*]
\item We present a decomposition-assembly framework for the top-k similarity search over knowledge graphs, which is the first work that considers the semantic-guided and response-time-bounded characteristics in one system.

\item We present an A* semantic search to find semantically similar graph matches, that can efficiently prune unpromising paths through a well-designed path semantic similarity ($pss$) estimation. We prove the effectiveness guarantee of our algorithm.


\item We optimize the A* semantic search to enable a trade-off between effectiveness and efficiency with a time bound $\mathcal{T}$. We prove that this method can converge to the globally optimal results as more time is given.

\item We evaluate the effectiveness and efficiency of our approach through extensive experiments on three real-world and large-scale knowledge graphs.

\end{itemize}

\vspace{-0.1cm}
\section{Related Work}
\label{previous_work}
According to how previous approaches process graph query, we categorize related work as follows.

\noindent\underline{\textbf{Graph pattern matching.}} Graph pattern matching is defined in terms of subgraph isomorphism \cite{Zou2011,Cheng2008}, which is NP-complete and is often too restrictive to capture sensible matches. Hence, graph simulation based pattern matching is proposed, such as \cite{Fan2010a,Ma2014}. These methods cannot be directly deployed to support graph query over knowledge graphs, because they do not consider the semantic constraints on edges even though they can map an edge to an $n$-hop path.

\noindent\underline{\textbf{Graph similarity search.}} Many efforts have been made for the graph similarity search based on different similarity metrics: (1) structural similarity search \cite{Khan2013,Khan2011,Jin2015}, (2) graph edit distance based search \cite{Zheng2013,Zeng2009}, and (3) weak semantic similarity search \cite{Yang2014,Yang2016,Zheng2016}. Note that, \cite{Khan2013,Jin2015} support edge-to-path mapping (however, do not consider the semantic constraints). Besides, \cite{Zheng2016} can find $n$-hop paths that are similar to a query edge based on prior knowledge.  Unlike \cite{Zheng2016}, we can find better $n$-hop paths (i.e., semantically similar) without external knowledge.


\noindent\underline{\textbf{Query-by-examples.}} Query-by-Example (QBE) aims to allow users to express their search intentions with examples. GQBE \cite{Jayaram2015} and Exemplar \cite{Mottin2016} are proposed for searching matches that are same as their counterparts from the examples. Moreover, \cite{Namaki2019,Song2019} are proposed to pose exemplars characterized by tuple patterns, and identify answers close to exemplar. Our approach can extend these QBE methods by returning more semantically similar answers to the given exemplar queries.

\noindent\underline{\textbf{Other methods to query knowledge graph.}} The knowledge graph search can also be conducted by the following query forms: (1) keywords search \cite{Namaki2018,Han2017}, (2) SPARQL search \cite{Peng2019,Zhang2018,Zou2011}, and (3) natural language search \cite{Hu2018,Zou2014,Zheng2018}. Most of these methods transform the input texts to query graphs for graph searching, so our graph query approach can be used to improve their performance.

\section{Preliminaries}
\label{pre_overview}
\vspace{-0.1cm}
\subsection{Background}
\label{pre}

\begin{myDef}
\label{def:kg}
\textbf{Knowledge graph}. A knowledge graph is defined as a graph $G=(V,E,L)$, with the node set $V$, edge set $E$, and a label function $L$, where (1) each node $u\in V$ represents an entity, (2) $E$ is an ordered subset of $V\times V$, each directed edge $e=u_iu_j\in E$ denotes the relationship between two entities $u_i$ and $u_j$, and (3) $L$ assigns a name and various types on each node, and a predicate on each edge.
\end{myDef}
\vspace{0.1cm}

Similar to \cite{Song2019,Ma2019,Yang2016}, we define the type as a label on each entity, rather than a node connecting to an entity with the predicate \textit{isA}. This assumption can reduce the size ($|E|$) of $G$, it also avoids dealing with irrelevant edges \textit{isA}.

\begin{myExample}
We assume that each node $u$ in a knowledge graph $G$ has at least one type and a unique name \cite{Zheng2016, Zheng2018}, e.g., $L(u).type=\{$\textit{Automobile}$\}$ and $L(u).name=$\textit{Audi\_TT}. For each edge $e$, we assign a predicate as $L(e)=$\textit{assembly}. If the type of a node in $G$ is unknown, we employ a probabilistic model-based entity typing method to assign a type on it \cite{Nakashole2013}.
\end{myExample}

\vspace{0.1cm}
\begin{myDef}
\label{def:qg}
\textbf{Query graph}. A query graph is defined as a graph $G_Q=(V_Q,E_Q,L_Q)$, with query node set $V_Q$, edge set $E_Q$, and a label function $L_Q$, where (1) $V_Q=V^s\cup V^t$, (2) $V^s=\{v^s\}$ is a set of \textit{specific nodes}, both the type and name of $v^s$ are known, (3) $V^t=\{v^t\}$ refers to \textit{target nodes}, only the type of $v^t$ is known, (4) $\forall e\in E_Q$ has a predicate $L_Q(e)$.
\end{myDef}
\vspace{0.1cm}

Since we assume that users do not have full knowledge about the dataset, so they are allowed to represent the query nodes without using the controlled vocabulary. For the example query graph in Figure \ref{fig:fullexample}, $V^s=\{v_3\}$ (type: \textit{Country} and name: \textit{Germany}), $V^t=\{v_1,v_2\}$ (respective types: \textit{Car}, \textit{Person}), and \textit{Car} is not a term defined in DBpedia's vocabulary.

\vspace{0.1cm}
\noindent\underline{\textbf{Query graph decomposition}}. We adopt a decomposition-assembly framework for $G_Q$. We decompose $G_Q$ into sub-query graphs for querying, and then assemble the matches of all sub-query graphs to form the top-k matches of $G_Q$.

\vspace{0.1cm}
\begin{myDef}
\label{def:subqg}
\textbf{Sub-query graph}. We define a sub-query graph of $G_Q$ as a graph $g_i=(V_i,E_i,L_Q)$, with the query node set $V_i$, query edge set $E_i$, and the same label function $L_Q$ as in $G_Q$, where (1) $g_i$ is a path from a specific node $v^s$ to a target node $v^t$, denoted by $\overline{v^sv^t}$, (2) and $V_Q=\cup V_i$ and $E_Q=\cup E_i$ over all sub-query graphs $g_i$ of $G_Q$.
\end{myDef}
\vspace{0.1cm}

\begin{myExample}
\label{example:subquery}
The query graph in Figure \ref{fig:fullexample} can be decomposed into two sub-query graphs: (1) find automobiles produced in Germany, denoted as $g_1$: $\langle v_1$-\textit{product}-$v_3\rangle$, and (2) find automobiles designed by Germans, denoted as $g_2$: $\langle v_1$-\textit{designer}-$v_2$-\textit{nationality}-$v_3\rangle$.
\end{myExample}
\vspace{0.1cm}

In general, all sub-query graphs intersect at a target node (called \textit{pivot node} $v^p$), e.g., $v_1$ in Example \ref{example:subquery}. Therefore, we can assemble the final matches via a join operation at $v^p$. The objective of query graph decomposition is to derive a number of sub-query graphs with an appropriate pivot node, to minimize the cost of query processing (Eq. \ref{eq:decomposition}). We use the possible search space as the cost metric (similar to \cite{Yang2016}) and resolve this problem through dynamic programming. The time complexity is $O(|V_Q|\cdot |E_Q|^2)$ according to \cite{Yang2016}. We show the impact of query graph decomposition on performance and its scalability in Section \ref{effective} and Section \ref{pivot}, respectively.

\vspace{-0.3cm}
\begin{equation}
\label{eq:decomposition}
\mathop{\arg\min}_{\{g_1...g_n\}} \ \sum_{i=1}^n cost(g_i)
\end{equation}
\vspace{-0.3cm}

According to \cite{Bonifati2017}, the chain, star, tree, cycle, and flower are common query graph shapes in knowledge graph search, so we provide an analysis on the effect of these query graph shapes with different sizes in Section \ref{pivot}.

\vspace{0.1cm}
\noindent\underline{\textbf{Sub-query graph matching}}. For each sub-query graph, we aim to find the semantically similar matches by identifying the candidate node (edge) matches for each query node (edge).

\vspace{0.1cm}
\noindent(1) \textit{Node match}. To overcome the mismatch in query nodes, we define a one-to-many relation $\phi$: $V_i\rightarrow V$ considering three cases: \textit{Identical}, \textit{Synonym}, and \textit{Abbreviation}. For each query node $v\in V_i$, $\phi(v)=\{u_1...u_n\}$ is a set of candidate matches in $V$, where the type (name) of $v$ is the same, synonymous, or abbreviated for the type (name) of $u$.

Usually, each node $u$ in a knowledge graph $G$ can have multiple types \cite{Nakashole2013}, then $u$ is a node match of the query node $v$ if one of $u$'s types satisfies the relation $\phi$. Although we assume that each query node $v$ has a user-specific type in Definition \ref{def:qg}, we still need to consider the following cases to enhance the robustness. (1) If a query node $v$ has multiple types, then we separately consider each type of $v$ in the relation $\phi$ for node matching, and then consider union of all of them as the overall node matches for $v$ . (2) If a user provides a query node $v$ without a type, then $v$ is a wildcard query node and can be mapped to the nodes with different types in a knowledge graph. In Section \ref{weight}, we introduce how to implement the relation $\phi$ by building a transformation library.

\vspace{0.1cm}
\noindent(2) \textit{Edge match}. To overcome the mismatch in query edges, we support the semantically edge-to-path mapping. Given a sub-query graph $g_i$ and a knowledge graph $G$, a path $\overline{u_iu_j}\in G$ is a match of an edge $v_iv_j\in g_i$, if $u_i\in \phi(v_i)$ and $u_j\in \phi(v_j)$. While considering paths, we ignore edge directionalities. Besides, we expect that the path $\overline{u_iu_j}$ is semantically similar to the edge $v_iv_j$. We elaborate this part in Defintion \ref{def:pss}.

Considering the sub-query graph $g_1$ in Figure \ref{fig:fullexample}, the edge matches of \textit{product} are the paths from \textit{Germany} to entities with type \textit{Automobile}, e.g., ($e_1$, $e_2$), ($e_9, e_{10}, e_{11}$), etc. We need to identify the most semantically similar path ($e_1$, $e_2$) from other edge matches, which motivates us to define the semantic graph $SG_Q$ for each sub-query graph $g_i$ as follows.

\vspace{0.1cm}
\begin{myDef}
\label{def:sg}
\textbf{Semantic graph}. Given a sub-query graph $g_i=(V_i,E_i,L_Q)$ and a knowledge graph $G=(V,E,L)$, the semantic graph is a weighted sub-graph of $G$ defined as $SG_Q=(V',E',L,W)$, with the node set $V'\subseteq V$, edge set $E'\subseteq E$, and weight set $W$, where (1) for each node $v\in V_i$, its node match $u\in \phi(v)$ belongs to $V'$, (2) for each edge $e=v_iv_j\in E_i$, its edge match $\overline{u_iu_j}\in SG_Q$ (i.e., $\forall e'\in\overline{u_iu_j}$ belongs to $E'$), (3) each $e'$ has a weight $w\in W$ to represent the semantic similarity between $e'$ and $e$ (Section \ref{embedding}).
\end{myDef}
\vspace{0.1cm}

According to Definition \ref{def:subqg}, a sub-query graph $g_i$ is a path graph denoted as $\overline{v^sv^t}$. So, we define the match of $g_i$ as a path in $SG_Q$ that is semantically similar to $\overline{v^sv^t}$.

\vspace{0.1cm}
\begin{myDef}
\label{def:submatch}
\textbf{Sub-query graph match}. Given a sub-query graph $g_i=\overline{v^sv^t}$ and a semantic graph $SG_Q$, a path $\overline{u^su^t}\in SG_Q$ is a match of $g_i$ if (1) $\overline{u^su^t}$ comprises the edge match of each edge $v_iv_j\in g_i$, (2) the path semantic similarity ($pss$) of $\overline{u^su^t}$ to $\overline{v^sv^t}$ equals or is greater than a predefined threshold $\tau$, denoted by $\psi(\overline{u^su^t},\overline{v^sv^t})\geq \tau$.
\end{myDef}
\vspace{0.1cm}

\begin{myDef}
\label{def:pss}
\textbf{Path semantic similarity ($pss$)}. We define the $pss$ $\psi(\overline{u^su^t},\overline{v^sv^t})$ as a function $f(w_1...w_n)$ of all weights appearing in match $\overline{u^su^t}$, which measures the semantic similarity of $\overline{u^su^t}$ to $g_i$. The details are given in Section \ref{pss}.
\end{myDef}
\vspace{0.1cm}

\vspace{0.1cm}
\noindent\underline{\textbf{Assembly}}. For each sub-query graph $g_i$, we can obtain a set of sub-query graph matches, denoted by $M_i=\{\overline{u^su^t}\}$. The sub-query graph matches from different $M_i$ may intersect at the same pivot node match $u^p\in\phi(v^p)$, so we can assemble them at $u^p$ to form a match for the query graph $G_Q$. Figure \ref{fig:fullexample} shows the assembly procedure. First, we find some matches with the greatest $pss$ for the two sub-query graphs $g_1$ and $g_2$. The match $\langle$\textit{Germany}-$e_1$-$e_2$-\textit{Audi\_TT}$\rangle$ from $M_1$ and $\langle$\textit{Germany}-$e_3$-$e_4$-\textit{Audi\_TT}$\rangle$ from $M_2$ can be assembled at pivot node match \textit{Audi\_TT} to form a match for $G_Q$. We define the match score of a match for $G_Q$ as the sum of $pss$ for all involved sub-query graph matches that intersect at the same $u^p$.

\vspace{-0.2cm}
\begin{equation}
\label{eq:ms}
S_m(u^p)={\displaystyle \sum\limits_{M_i}\psi(\overline{u^su^t},\overline{v^sv^t})}
\end{equation}
\vspace{-0.3cm}
\begin{equation}\nonumber
\label{eq:ms1}
s.t. \quad \overline{u^su^t}\in M_i\text{ that contains the same } u^p
\end{equation}
\vspace{-0.5cm}

The best match of $G_Q$ is the one with the greatest match score, such as the top-1 match involving \textit{Audi\_TT} in Figure \ref{fig:fullexample} (with the greatest match score 1.891=0.891+1).

\vspace{-0.1cm}
\subsection{Problems}
\label{problem}
According to the background above, we derive two major problems that need to be resolved in this paper as follows.

\vspace{0.1cm}
\noindent\underline{\textbf{Problem 1}}. \textit{Given a query graph $G_Q=\{g_1...g_n\}$ and a knowledge graph $G$, we find the top-k matches $M$ according to the match score $S_m(u^p)$  as follows.}

\vspace{-0.3cm}
\begin{equation}
\label{eq:problem1.1}
M = \sigma_{\max(S_m)}(\Join_{u^p}M_i)
\end{equation}
\begin{equation}\nonumber
\label{eq:problem1}
s.t.\quad |M|=k,\quad M_i = \{\mathop{\arg\max}_{\overline{u^su^t}} \ \psi(\overline{u^su^t},\overline{v^sv^t})\}
\end{equation}
\vspace{-0.3cm}

In Eq. \ref{eq:problem1.1}, we use $\Join_{u^p}$ to denote the assembly at $u^p$ and use $\sigma_{\max(S_m)}$ to denote the top-k matches selection based on the match score $S_m(u^p)$. $M_i=\{\overline{u^su^t}\}$ are the sub-query matches with the greatest $pss$ for each $g_i$. This problem is non-trivial, because (1) we need to find globally optimal $M_i$ for each $g_i$, and (2) the assembly is computationally expensive if $G_Q$ has a large number of $g_i$, and each one has many candidate matches. We solve this problem
efficiently in \textbf{Section \ref{Asearch}}.

\vspace{0.1cm}
\noindent\underline{\textbf{Problem 2}}. \textit{Given a query graph $G_Q=\{g_1...g_n\}$ and a knowledge graph $G$, we find the approximate top-k matches $\hat{M}$ within a user-specified time bound $\mathcal{T}$ as follows.}

\vspace{-0.3cm}
\begin{equation}
\label{eq:problem2}
\hat{M} = \sigma_{\max(S_m)}(\Join_{u^p}\hat{M}_i)
\end{equation}
\begin{equation}\nonumber
\label{eq:problem2.1}
s.t.\ |\hat{M}|=k,\ \ \text{time bound}\ \mathcal{T},\ \left (\frac{\hat{M}\cap M}{\hat{M}\cup M}\right )_\mathcal{T'}\geq \left (\frac{\hat{M}\cap M}{\hat{M}\cup M}\right
)_\mathcal{T}
\end{equation}
\vspace{-0.3cm}

We use the Jaccard similarity of $\hat{M}$ and $M$ to measure the degree of approximation. With more time given ($\mathcal{T'}>\mathcal{T}$), $\hat{M}$ can approach $M$. The key of this problem is how to return $\hat{M}$ quickly, and refine it as more time is given. Moreover, we need to prove that the globally optimal results can be returned if sufficient time is given (e.g., $\hat{M}=M$). We deal with this problem in \textbf{Section \ref{bound}}.

\section{Semantic Graph Construction}
\label{sg}
In this section, we leverage a knowledge graph embedding model to construct the semantic graph $SG_Q$, then present the predicate semantic similarity ($pss$) based on $SG_Q$.

\vspace{-0.1cm}
\subsection{Knowledge Graph Embedding}
\label{embedding}
Knowledge graph embedding aims to represent each predicate and entity of a knowledge graph $G$ as an $n$-dimensional semantic vector, such that the original structures and relations in $G$ are preserved in these learned semantic vectors \cite{Huang2019}. We summarize the core idea of most existing knowledge graph embedding methods  as follows: (1) initialize the vector of each element in triple $\langle$\textit{h},\textit{r},\textit{t}$\rangle$ as $\langle\bm{h}$,$\bm{r}$,$\bm{t}\rangle$, where \textit{h/t} indicates the head/tail entity and \textit{r} denotes the predicate, (2) define a function $g()$ to measure the relation of $\langle\bm{h}$,$\bm{r}$,$\bm{t}\rangle$, and optimize $g()$ to satisfy $\bm{t}\approx g(\bm{h},\bm{r})$. The predicate semantic space $\bm{E}=\{\bm{e_1}...\bm{e_n}\}$ is an output of a knowledge graph embedding model. The semantic similarity between the two edges can be represented by the similarity between two predicate vectors.

In this paper, we use the cosine similarity to measure the similarity between two predicate vectors. Each weight $w$ in the semantic graph $SG_Q$, denoted by $sim(L_Q(e),L(e'))$, e.g., $sim($\textit{product}, \textit{assembly}$)$, can be calculated as follows.

\vspace{-0.3cm}
\begin{equation}
\label{eq:cosine}
w=sim(L_Q(e),L(e'))=\frac{\bm{e}\cdot \bm{e'}}{||\bm{e}||\times||\bm{e'}||}
\end{equation}
\vspace{-0.3cm}

We next introduce how to preserve these weights on a knowledge graph to generate semantic graph.

\setlength{\textfloatsep}{0.2cm}
\begin{table}
\setstretch{1.0}
\setlength{\abovecaptionskip}{0.1cm}
 \fontsize{8pt}{3.3mm}\selectfont
  \centering
  \caption{Transformation library}
  \begin{tabular} {c|c}
   \hline
   \textbf{\textit{Synonym} and \textit{abbreviation} records} & \textbf{Types and names} \\
   \hline
   \textit{Car}, \textit{Motorcar}, \textit{Auto}, \textit{Vehicle} & type: $<$\textit{Automobile}$>$ \\
   \hline
   \textit{GER}, \textit{FRG}, \textit{Federal Republic of Germany} & name: \textit{Germany} \\
   \hline
  \end{tabular}
  \label{tab:library}
\end{table}

\subsection{Constructing Semantic Graph}
\label{weight}
\begin{figure}
\setlength{\abovecaptionskip}{0.1cm}
\setlength{\belowcaptionskip}{-0.2cm}
\centerline{\includegraphics[scale=0.47]{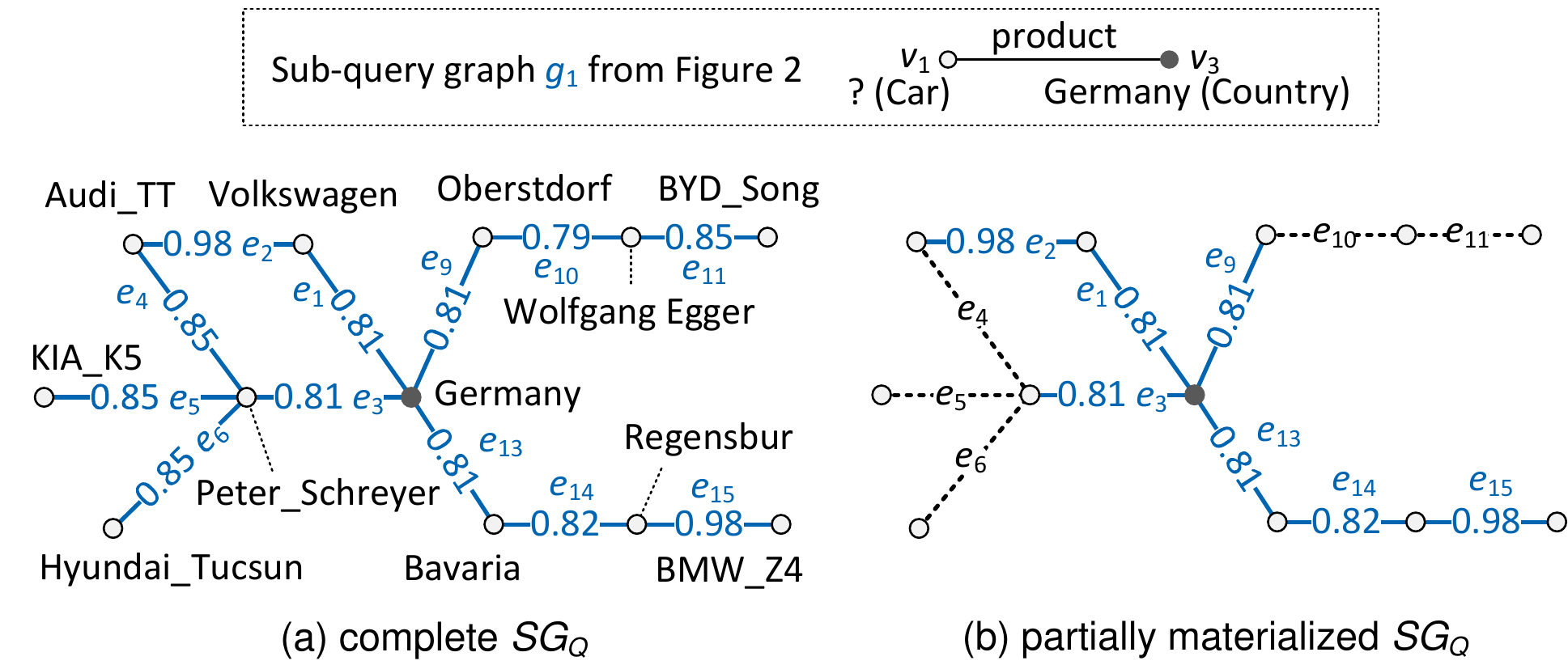}}
\caption{Semantic graph construction: all predicates in the semantic graph are provided as \textit{assembly}:$\{e_2,e_{15}\}$, \textit{country}:$\{e_1,e_9,e_{13}\}$, \textit{federalState}: $\{e_{14}\}$, \textit{nationality}: $\{e_3\}$, \textit{designer}:$\{e_4,e_5,e_6,e_{11}\}$, and \textit{birthPlace}: $\{e_{10}\}$.}
\label{fig:semanticgraph}
\end{figure}

Figure \ref{fig:semanticgraph}(a) shows an example semantic graph $SG_Q$ for the sub-query graph $g_1$ in Figure \ref{fig:fullexample}. A straightforward idea to build $SG_Q$ is: (1) find the node matches of each query node, e.g., $\phi(v_1)$=$\{$\textit{Audi\_TT, KIA\_K5, BYD\_Song, Hyundai\_Tucsun, BMW\_Z4}$\}$, and $\phi(v_3)$=\textit{Germany}, (2) find the edge matches of each query edge, e.g., edge matches of \textit{product} include paths $(e_1,e_2)$, $(e_3,e_4)$, $(e_3,e_5)$, $(e_3,e_6)$, $(e_9,e_{10},e_{11})$, and $(e_{13},e_{14},e_{15})$, and (3) assign weights on edges through Eq. \ref{eq:cosine}.

\vspace{0.1cm}
\noindent\underline{\textbf{Analysis}}. To find all edge matches for a query edge $v_iv_j\in g_i$, we must enumerate all possible paths between $u_i\in \phi(v_i)$ and $u_j\in \phi(v_j)$. However, the high connectivity of a knowledge graph $G$ makes it computationally expensive.

\vspace{0.1cm}
\noindent\underline{\textbf{A lightweight way}}. Figure \ref{fig:semanticgraph}(b) shows an alternative way to construct $SG_Q$ on the fly. We \textit{push down} the semantic graph construction to the query processing stage, which means that $SG_Q$ is partially materialized (not completely constructed in advance). Given a sub-query graph $g_i=\overline{v^sv^t}$, we materialize the $SG_Q$ as follows.


\vspace{0.1cm}
\noindent\underline{(1) Get node matches of $v^s$}. To implement the relation $\phi$ for node matching, we build a \textit{synonym} and \textit{abbreviation} transformation library \cite{Yang2014} for all types and names existing in $G$ based on BabelNet (the largest multilingual synonym dictionary \cite{Navigli2010}). For instance, we invoke the API of BabelNet to collect the synonyms and abbreviations of the input keyword as one row in Table \ref{tab:library}. For each query node $v^s$, we use this library to find its node matches $\phi(v^s)$ through the \textit{synonym} or \textit{abbreviation} transformation, e.g., a query node with type \textit{Car} is mapped to a set of entities with type \textit{Automobile} in $G$.

\vspace{0.1cm}
\noindent\underline{(2) Materialize the 1-hop $SG_Q$ for $u^s\in\phi(v^s)$}. Given a node match $u^s\in \phi(v^s)$, we assign the weight $w$ on each edge that connects to $u^s$ based on Eq. \ref{eq:cosine}, generating a 1-hop $SG_Q$ for $u^s$. For example, weighted edges $\{e_1,e_3,e_9,e_{13}\}$ in Figure \ref{fig:semanticgraph}(b) act as the 1-hop $SG_Q$ for the node \textit{Germany}.

\vspace{0.1cm}
\noindent\underline{(3) Next-hop decision}. Given a partially materialized $SG_Q$, we select a next-hop node for further querying. The selected node should be the one with the greatest probability of finding the best match for $g_i$. We show the details in Section \ref{Asearch}.

\vspace{0.1cm}
\noindent\underline{(4) Termination check}. Starting from the next-hop, we repeat the above steps to materialize $SG_Q$ gradually, until a node match $u^t\in \phi(v^t)$ is detected. The path $\overline{u^su^t}$ is a match of $g_i$. In Figure \ref{fig:semanticgraph}(b), we can find the best matches from the partially materialized $SG_Q$ (blue lines) excluding the dashed lines.

\vspace{0.1cm}
\noindent\underline{\textbf{Remarks}}. To construct a complete $SG_Q$, we need $O(|E'|)$ time to assign weights on all the edges of $SG_Q$. The time complexity is expected to be reduced by constructing $SG_Q$ partially, because a good next-hop selection would reduce the size of $SG_Q$, pruning the search space significantly. Moreover, a good next-hop selection also ensures that the sub-query graph match with the greatest path semantic similarity ($pss$) can be found. We show the details of $pss$ in Section \ref{pss} and show the semantic-guided search based on $pss$ in Section \ref{Asearch}.

\vspace{-0.2cm}
\subsection{Path Semantic Similarity}
\label{pss}
We define the path semantic similarity ($pss$) of a sub-query graph match $\overline{u^su^t}$ to $g_i$ based on the following observations.
\begin{itemize}[leftmargin=*]
\item A match $\overline{u^su^t}$ comprises a set of edges $\{e_1...e_n\}$. Each edge $e_i$ has a weight $w$ that indicates the semantic similarity to one edge $e\in g_i$. Hence, the $pss$ should be a function of all weights appearing at $\overline{u^su^t}$.
\item According to Eq. \ref{eq:cosine}, two edges are semantically similar if their predicate vectors are similar. Thus, the edges with greater $w$ would be more beneficial to the $pss$.
\item A smaller $w$ usually indicates that two edges show different semantic meanings. Therefore, the edges with smaller $w$ have a negative effect on the $pss$.
\end{itemize}

\begin{myExample}
In Figure \ref{fig:semanticgraph}(a), the paths $(e_1,e_2)$ and $(e_{13},e_{14},e_{15})$ are more semantically similar to $g_1$ than others, because the predicate vectors of edges $e_2$ and $e_{15}$ (\textit{assembly}) are more similar to the one of edge \textit{product} (with the greatest $w$=0.98). And other paths containing edges such as $e_4$ (\textit{designer}), $e_{10}$ (\textit{birthPlace}), etc., show the different meanings to $g_1$, because $e_4$ and $e_{10}$ are less semantically similar to \textit{product}.
\end{myExample}
\vspace{0.1cm}

Based on these intuitions, we calculate the $pss$ of the match $\overline{u^su^t}$ to $\overline{v^sv^t}$, denoted by $\psi(\overline{u^su^t},\overline{v^sv^t})$, as the geometric mean of all weights appearing at the match $\overline{u^su^t}$.

\vspace{-0.2cm}
\begin{equation}
\label{eq:pss}
\psi(\overline{u^su^t},\overline{v^sv^t})=\sqrt[n]{\prod_{\forall w_j\in \overline{u^su^t}}{w_j}}
\end{equation}
\vspace{-0.3cm}

\section{Semantic-guided Search}
\label{Asearch}

In this section, we first present an A* semantic search to find the top-k matches from $SG_Q$ with the greatest path semantic similarity ($pss$) for each sub-query graph $g_i\in G_Q$. We then assemble all matches for $g_i$ to form the final matches for $G_Q$. The classic A* search \cite{Hart1968} finds the shortest path based on a heuristic length estimation. Here, we design the \textit{pss} estimation based on semantics to find the most semantically similar paths. The basic idea of A* semantic search is that we compute a heuristic $pss$ estimation for a possible match at each detected node, and gradually expand the search space following the guidance of the estimated $pss$ until a match with maximum $pss$ is found. We achieve two benefits from a good $pss$ estimation: (1) we can find the globally optimal matches of $g_i$, and (2) we can prune the search space significantly.

We next introduce the heuristic $pss$ estimation in Section \ref{estimate}, then discuss A* semantic search based on $pss$ estimation and prove the effectiveness guarantee in Section \ref{Asearch_overview}. Finally, we show the assembling of the matches in Section \ref{final}.

\subsection{Heuristic Estimation of $pss$}
\label{estimate}
Given a match $\overline{u^su^t}$ of a sub-query graph $g_i$, it can be divided into an explored partial path $\overline{u^su_i}$ and an unexplored partial path $\overline{u_iu^t}$ at each detected node $u_i$. We compute the upper bound of the exact $pss$ $\psi(\overline{u^su^t},\overline{v^sv^t})$ at $u_i$ as the estimated $pss$, denoted by $\hat{\psi}(\overline{u^su_i},\overline{v^sv^t})$ ($\hat{\psi}_i$ for short), by considering the semantic information from both partial paths. Based on the estimated $pss$, we can effectively prune the search space and find the sub-query graph match with the greatest $pss$ due to the following reasons. (1) Suppose that we already find a sub-query graph match, then we can safely prune the potential matches having the smaller $\hat{\psi}_i$ than the explored match's exact $pss$. Only the potential matches with a greater $\hat{\psi}_i$ than the explored match's exact $pss$ would be considered for further searching. (2) Given a predefined $pss$ threshold $\tau$, we can prune the unpromising potential matches that have the $\hat{\psi}_i< \tau$ (we set $\tau=0.8$ in Section \ref{exp}).

We next introduce how to obtain the upper bound $\hat{\psi}_i$ of $\psi$ for the $pss$ estimation. And we prove the effectiveness guarantee in Section \ref{Asearch_overview} (Theorem \ref{th:2}), that is, the match with the greatest $pss$ must be found.

According to Eq. \ref{eq:pss}, we need the weight product ($\prod w_j$) and the path length ($n$) of a match $\overline{u^su^t}$ to compute the exact $\psi$. So we first estimate the upper bound of the weight product and path length, in order to estimate the upper bound of $\psi$.

\vspace{0.1cm}
\noindent\underline{\textbf{Upper bound of the weight product}}. The weight product of $\overline{u^su^t}$ is divided into two parts at each detected node $u_i$.

\begin{itemize}[leftmargin=*]
\item The weight product of partial path $\overline{u^su_i}$, e.g., $\prod_{\forall w_j\in \overline{u^su_i}}w_j$.
\item The weight product of partial path $\overline{u_iu^t}$, e.g., $\prod_{\forall w_j\in \overline{u_iu^t}}w_j$.
\end{itemize}

We can compute the exact weight product of $\overline{u^su_i}$ because $\overline{u^su_i}$ is explored in the partially materialized $SG_Q$. On the other hand, the partial path $\overline{u_iu^t}$ is unexplored, so we use the maximum weight of all adjacent edges of $u_i$ as the upper bound of the weight product of $\overline{u_iu^t}$, denoted as $m(u_i)$.

\vspace{0.1cm}
\begin{myLemma}
\label{lemma:weightbound}
The maximum weight $m(u_i)$ is the upper bound of the weight product of the partial path $\overline{u_iu^t}$.
\end{myLemma}
\vspace{0.1cm}

\begin{IEEEproof}
Given the weight product $\prod_{w_j\in \overline{u_iu^t}}w_j$ of $\overline{u_iu^t}$, $w_j$ indicates the $j$-th weight in $\overline{u_iu^t}$. Due to the monotonicity of weight product, we have $w_1\geq \prod_{w_j\in \overline{u_iu^t}}w_j$. Also, $m(u_i)\geq w_1$ because we assume $m(u_i)$ is the max weight of all adjacent edges of $u_i$. Hence, we have $m(u_i)\geq \prod_{w_j\in \overline{u_iu^t}}w_j$.
\end{IEEEproof}
\vspace{0.1cm}

\noindent\underline{\textbf{Upper bound of the path length}}. Since different matches have different path lengths, it is difficult to get a uniform upper bound of the path length $n$ for all matches. Hence, we relax the upper bound of the exact path length to the upper bound of the user desired path length. If a user wants to find the top-k matches within $\hat{n}$-hop, then only the matches having $n\leq \hat{n}$ ($\hat{n}$-bounded match) will be returned. Hence, $\hat{n}$ is the upper bound of the path length for all the $\hat{n}$-bounded matches.

\vspace{0.1cm}
\noindent\underline{\textbf{Estimated $pss$ of $\hat{n}$-bounded match}}. Given the above two upper bounds. We compute the estimated $pss$ $\hat{\psi}_i$ at each node $u_i\neq u^t$ as follows, where $u^t\in\phi(v^t)$ is a target node match. And we set $\hat{\psi}_i$ equals to the exact $pss$ $\psi$ when $u_i=u^t$.

\vspace{-0.2cm}
\begin{equation}
\label{eq:estimatedPss}
\hat{\psi}(\overline{u^su_i},\overline{v^sv^t})=\sqrt[\hat{n}]{\prod_{\forall w_j\in \overline{u^su_i}}w_j\cdot m(u_i)}
\end{equation}
\vspace{-0.2cm}

\begin{myTheorem}
\label{th:1}
The $pss$ estimation $\hat{\psi}_i$ is the upper bound of the exact $pss$ $\psi$ of the match $\overline{u^su^t}$=$\overline{u^su_i}$+$\overline{u_iu^t}$ with the path length $n\leq \hat{n}$, where $\hat{n}$ is the user desired path length.
\end{myTheorem}

\vspace{0.1cm}
\begin{IEEEproof}
We use notation $W_{si}$ ($W_{it}$) to denote the weight product of the partial path $\overline{u^su_i}$ ($\overline{u_iu^t}$). If $u_i$$\neq$$u^t$, then $\hat{\psi}_i$=$\sqrt[\hat{n}]{W_{si}\cdot m(u_i)}$$\geq$$\sqrt[n]{W_{si}\cdot m(u_i)}$, because $\hat{n}$$\geq$$n$ and the $n$-th root $W_{si}$$\cdot$$m(u_i)$$\in$(0,1]. Moreover, we have $m(u_i)$$\geq$$W_{it}$ based on Lemma \ref{lemma:weightbound}, so that $\sqrt[n]{W_{si}\cdot m(u_i)}$$\geq$$\sqrt[n]{W_{si}\cdot W_{it}}$=$\psi$. Hence, $\hat{\psi}_i$$\geq$$\psi$ holds. On the other hand, if $u_i$=$u^t$, then $\hat{\psi}_i$=$\psi$. In summary, $\hat{\psi}_i$$\geq$$\psi$ holds for all cases.
\end{IEEEproof}
\vspace{0.1cm}

\noindent\underline{\textbf{Remarks}}. (1) The user desired path length $\hat{n}$ is specified by users before graph querying. (2) Our A* semantic search can find the globally optimal $\hat{n}$-bounded matches (proved later in Theorem \ref{th:2}) based on the above heuristic $pss$ estimation.


\setlength{\textfloatsep}{0cm}
\begin{algorithm}[t]
\setstretch{0.8}
\small
\fontsize{7.8pt}{3.5mm}\selectfont
\caption{A* semantic search}
\label{alg:Asearch}
\KwData{sub-query graph $g_i$, number of matches $k$}
\KwResult{match set $M_i$}
$\forall u^s\in \phi(v^s)$:  $q$=$\{\langle u^s,\hat{\psi}_s\rangle\}$, \textit{visited}=$\{u_s\}$, $M_i$=$\emptyset$\;
\While{$q\neq \emptyset$}{
	$\langle\overline{u^su_i},\hat{\psi}_i\rangle$=$q$.pop\_max() \tcp*{\footnotesize{Next-hop selection}}
	\If(\tcp*[f]{\footnotesize{Search space expansion}}){$u_i\not\in\phi(v^t)$}{
		\For{$\forall u_l\in N(u_i)$}{
			\If{$!$visited.contains$(u_l)$}{
                \textit{visited}.add($u_l$)\;
				$\overline{u^su_l}$=$\overline{u^su_i}$+$u_iu_l$\;
				$\langle\overline{u^su_l},\hat{\psi}_l\rangle$=pssEstimation()\;
				\If{$\hat{\psi}_l\geq \tau$}{
					$q$.push\_heap($\langle\overline{u^su_l},\hat{\psi}_l\rangle$)\;
				}
			}
		}
	}
	\Else{
		$M_i$.push\_heap($\langle\overline{u^su_i},\hat{\psi}_i\rangle$)\;
		\If(\tcp*[f]{\footnotesize{Top-k matches check}}){$|M_i|=k$}{
			break\; 
		}
	}
}
\Return $M_i$\;
\end{algorithm}

\vspace{-0.1cm}
\subsection{A* Semantic Search}
\label{Asearch_overview}
In this section, we introduce our A* semantic search based on the above $pss$ estimation, illustrated in Algorithm \ref{alg:Asearch}.

\vspace{0.1cm}
\noindent\underline{\textbf{Notations}}. We use a max-heap as the match set $M_i$ for a sub-query graph $g_i$, to record each found match and its $pss$, e.g., $\langle\overline{u^su^t},\psi\rangle$. We use another max-heap as the priority queue $q$ to record each explored partial path $\overline{u^su_i}$ and its estimated $pss$, e.g., $\langle\overline{u^su_i},\hat{\psi}_i\rangle$. For $u_i=u^s\in\phi(v^s)$, $\overline{u^su_i}$ is the node $u^s$ itself. Each node $u_i$ indicates a next-hop choice for search space expansion. We also use a hash set \textit{visited} to record all visited nodes, avoiding duplicate access. A $pss$ threshold $\tau$ is used to prune the unpromising matches having $\hat{\psi}_i< \tau$.

\vspace{0.1cm}
\noindent\underline{\textbf{Overview}}. Given a sub-query graph $g_i$=$\overline{v^sv^t}$, we start with the node match $u^s\in\phi(v^s)$ for A* semantic search (line 1). The main procedures are: (1) \textit{Next-hop selection}. We select the node $u_i$ with the greatest $\hat{\psi}_i$ as the next-hop for search space expansion, from the priority queue $q$ (line 3). (2) \textit{Search space expansion}. Starting from $u_i$, we expand the search space as $\overline{u^su_l}$=$\overline{u^su_i}$+$u_iu_l$ for each neighbour node $u_l$ of $u_i$, and compute $\hat{\psi}_l$ for each new partial path $\overline{u^su_l}$ (lines 5-9). All these $\langle\overline{u^su_l}$,$\hat{\psi}_l\rangle$ pairs ($\hat{\psi}_l\geq\tau$) are stored in $q$ for further exploration (lines 10-11). (3) \textit{Top-k matches check}. We repeat (1) and (2) until a match $\overline{u^su_i}$ is popped from $q$, where $u_i\in\phi(v^t)$. We record it in match set $M_i$ and terminate the search until $k$ matches are found (lines 13-15).

\begin{figure}
\setlength{\abovecaptionskip}{0.1cm}
\setlength{\belowcaptionskip}{0.1cm}
\centerline{\includegraphics[scale=0.46]{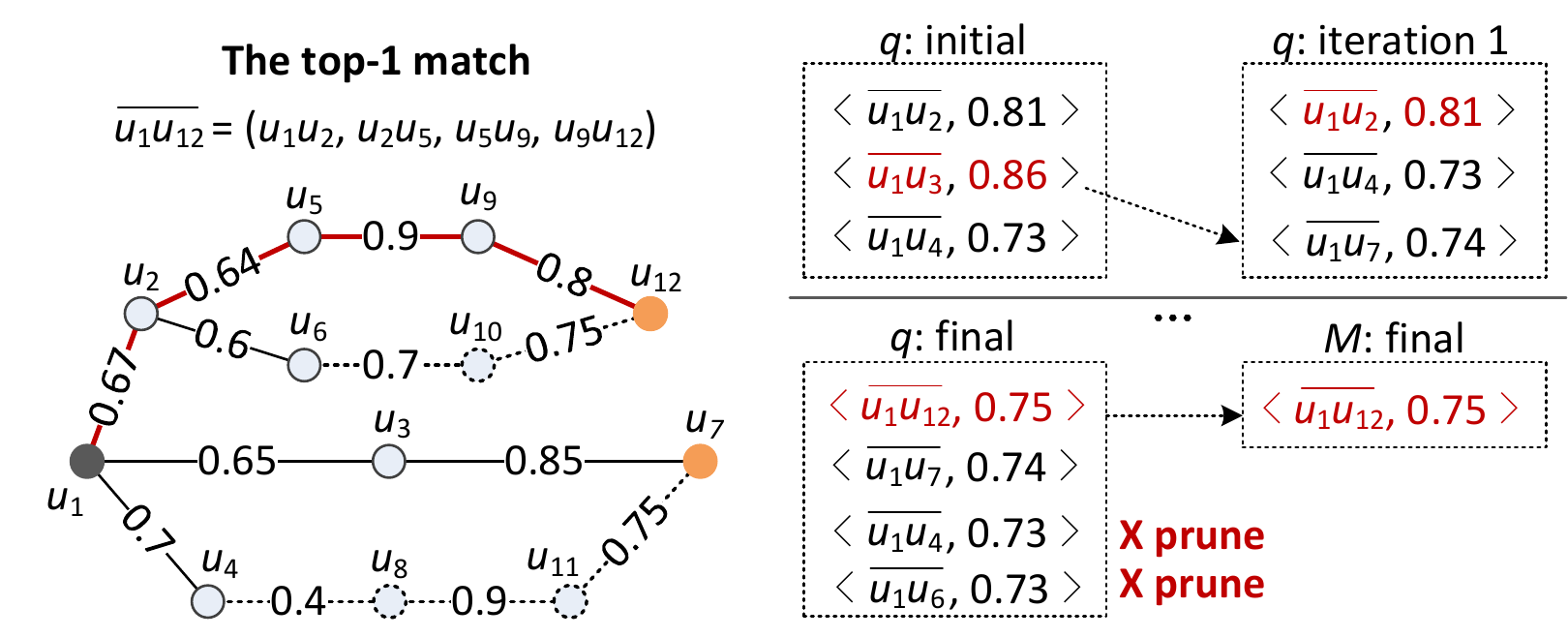}}
\caption{An example of the top-1 match searching}
\label{fig:astarexample}
\end{figure}

\vspace{0.1cm}
\begin{myExample}
\label{exp:casestudy1}
Figure \ref{fig:astarexample} shows an example for searching the top-1 match from a specific node match $u_1$ to target node matches $\{u_7$,$u_{12}\}$ (we set $\hat{n}$=4). The solid lines indicate the expanded search space, doted lines show the pruned data, and red lines denote the top-1 match. Finally, 38.5\% of edges and 25\% of nodes are pruned. At the beginning, we expand the search space from $u_1$, all its neighbours are added in the priority queue $q$, e.g., $q$=$\{\langle\overline{u_1u_2}$,0.81$\rangle$, $\langle\overline{u_1u_3}$,0.86$\rangle$, $\langle\overline{u_1u_4}$,0.73$\rangle\}$. We next select $u_3$ in $\overline{u_1u_3}$ to expand the search space because it has the greatest estimated $pss$ of 0.86, and we add the path $\overline{u_1u_7}$=$(u_1u_3,u_3u_7)$ in the priority queue $q$ as $\langle\overline{u_1u_7}$,0.74$\rangle$. However, we cannot return $\overline{u_1u_7}$ as the top-1 match because we still have $\langle\overline{u_1u_2},0.81\rangle$ in $q$. From $\overline{u_1u_2}$, we may find a better match with $pss$ of 0.81$>$0.74. Hence, we continue to expand the search space from $u_2$, $u_5$, $u_9$ until $u_{12}$ is detected. Finally, we have $q$=$\{\langle\overline{u_1u_{12}}$,0.75$\rangle$, $\langle\overline{u_1u_7}$,0.74$\rangle$, $\langle\overline{u_1u_4}$,0.73$\rangle$, $\langle\overline{u_1u_6}$,0.73$\rangle\}$, and $\overline{u_1u_{12}}$ is the top-1 match, while the potential matches from $\overline{u_1u_4}$ and $\overline{u_1u_6}$ can be safely pruned. This is because the upper bound of $pss$ for $\overline{u_1u_4}$ and $\overline{u_1u_6}$ are smaller than the $pss$ of the top-1 match.
\end{myExample}

\vspace{0.1cm}
\begin{myTheorem}
\label{th:2}
Our A* semantic search ensures that the sub-query graph match $\overline{u^su^t}$  with the greatest $pss$ can be found.
\end{myTheorem}
\vspace{0.1cm}

\begin{IEEEproof}
Suppose that the returned $\overline{u^su^t}$ is not the best match, then we have $\psi$$\leq$$\psi_{opt}$, where $\psi$ and $\psi_{opt}$ are the $pss$ of $\overline{u^su^t}$ and the best match, respectively. Since A* semantic search starts from $u^s$ and all its neighbours are considered in the priority queue $q$ for further expansion, then $q$ must contain one partial path $\langle\overline{u^su_i},\hat{\psi}_i\rangle$ that belongs to the best match. According to Theorem \ref{th:1}, we have $\hat{\psi}_i$$\geq$$\psi_{opt}$. Then we have $\hat{\psi}_i$$\geq$$\psi_{opt}$$\geq$$\psi$. Hence, our A* semantic search will continue to expand the search space from $u_i$ (line 3 in Algorithm \ref{alg:Asearch}) rather than returning the non-optimal match $\overline{u^su^t}$.
\end{IEEEproof}
\vspace{0.1cm}


\vspace{0.1cm}
\noindent\underline{\textbf{Complexity}}. The time consumption of our A* semantic search is dominated by the search space expansion. Given a partial path $\overline{u^su_i}$, we expand the search space from the node $u_i$ as follows: (1) construct a new partial path $\overline{u^su_l}$=$\overline{u^su_i}$+$u_iu_l$ for each neighbour node $u_l$ of $u_i$ and (2) update the priority queue $q$ with each $\langle\overline{u^su_l},\hat{\psi}_l\rangle$ pair. We use $V^*$ to denote all detected nodes in step (1), then the time complexity is $O(|V^*|\log |V^*|)$, where $O(\log |V^*|)$ is the time for the update of max-heap $q$.

\vspace{0.1cm}
\noindent\underline{\textbf{Remarks}}. (1) We implement the graph querying in a multithreaded manner (one thread for each $g_i\in G_Q$). (2) In general, we usually need more than $k$ matches collected for each $g_i$ to ensure that $k$ final matches can be assembled for $G_Q$.

\vspace{-0.1cm}
\subsection{Final Matches Assembly}
\label{final}
We employ the Threshold Algorithm (TA) \cite{Fagin2001} based method to efficiently assemble the top-k matches for $G_Q$.


\vspace{0.1cm}
\noindent\underline{\textbf{Core idea of assembly}}. Given the match sets $\{M_i\}$ for all sub-query graphs $\{g_i\}$, the TA-based assembly follows three steps. (1) It accesses all $M_i$ in descending order of the match's $pss$. Since $M_i$ is a max-heap, so we pop the best match in $M_i$ at each access. (2) It joins the detected matches with the same pivot node match $u^p$ to generate a final match $fm(u^p)$, and computes its upper and lower bound on match score, denoted by $\overline{S_m}(u^p)$ and $\underline{S_m}(u^p)$, respectively. (3) It terminates early if $k$ final matches are found, for which the smallest $\underline{S_m}(u^p)$ is larger than other final matches' greatest $\overline{S_m}(u^p)$.

We next introduce how to compute $\overline{S_m}(u^p)$ and $\underline{S_m}(u^p)$ in the TA-based assembly, then we prove that the top-k final matches can be returned through the TA-based assembly. According to Eq. \ref{eq:ms}, the match score $S_m(u^p)$ of a final match $fm(u^p)$ is computed by aggregating the $pss$ $\psi$ of the matches that contain the same pivot node match $u^p$, from each $M_i$. Intuitively, if we know the $pss$ bounds $\overline{\psi}$ and $\underline{\psi}$ of such a match from each $M_i$, then we can compute the bounds of match score during the TA-based assembly as follows.

\begin{equation}
\label{eq:bounds}
\overline{S_m}(u^p)=\sum_{M_i}\overline{\psi}\quad\text{ and }\quad\underline{S_m}(u^p)=\sum_{M_i}\underline{\psi}
\end{equation}
\vspace{-0.3cm}

\noindent where $\overline{\psi}$ ($\underline{\psi}$) is the upper (lower) bound on $pss$ of the match $\overline{u^su^t}\in M_i$ that contains the same pivot node match $u^p$.

\vspace{0.1cm}
\noindent\underline{\textbf{Upper bound of $S_m(u^p)$}}. At each access of the TA-based assembly, if the match that contains the same $u^p$ is not accessed from $M_i$ so far, then we have $\overline{\psi}=\psi_{cur}$, where $\psi_{cur}$ is the $pss$ of the current accessed match in $M_i$. This is because the TA-based assembly accesses each $M_i$ in the descending order of the match's $pss$. All the un-accessed matches from $M_i$ must have $\psi\leq\psi_{cur}$. If we find this match from $M_i$, then we have $\overline{\psi}=\psi$. The upper bound $\overline{\psi}$ is decreased as the TA-based assembly executes. Finally, $\overline{S_m}(u^p)$=$S_m(u^p)$, when the matches containing the same $u^p$ are accessed from each $M_i$.

\vspace{0.1cm}
\noindent\underline{\textbf{Lower bound of $S_m(u^p)$}}. At each access of the TA-based assembly, if the match that contains the same $u^p$ is not accessed from $M_i$ so far, then we have $\underline{\psi}=0$. This is because it is possible that $M_i$ does not contain such a match. If we find this match from $M_i$, then we have $\underline{\psi}=\psi$. The lower bound $\underline{\psi}$ is increased from 0 to $\psi$ as the TA-based assembly executes. Finally, $\underline{S_m}(u^p)=S_m(u^p)$, when the matches containing the same $u^p$ are accessed from each $M_i$.

\vspace{0.1cm}
\noindent\underline{\textbf{Termination check}}. We terminate the TA-based assembly if the top-k final matches are found. Specifically, (1) we sort the final matches in descending order of $\underline{S_m}(u^p)$, (2) we select the $k$-th largest $\underline{S_m}(u^p)$ as the lower bound of the top-k match score, denoted as $L$, (3) we select the greatest $\overline{S_m}(u^p)$ among other final matches as their upper bound on match score, denoted as $U$, and (4) we terminate the assembly if $L\geq U$.

\vspace{0.1cm}
\begin{myTheorem}
\label{th:assembly}
The TA-based assembly can obtain the top-k final matches, when $L\geq U$ holds.
\end{myTheorem}

\vspace{0.1cm}
\begin{IEEEproof}
Since the lower bound $\underline{\psi}$ increases from 0 to the exact $pss$ $\psi$ as the TA-based assembly processes, the lower bound of the top-k match score $L$ is also increased (Eq. \ref{eq:bounds}). Similar to $L$, $U$ is decreased because the upper bound $\overline{\psi}$ decreases from $\psi_{cur}$ to the exact $pss$ $\psi$ as the TA-based assembly processes. Hence, if $L\geq U$ holds at the $r$-th access of the TA-based assembly, then it will hold for all $r'>r$ accesses. Therefore, the TA-based assembly can terminate safely and return the top-k final matches when $L\geq U$ holds.
\end{IEEEproof}

\vspace{0.1cm}
\noindent\underline{\textbf{Complexity}}. In the worst case, all the matches from each match set $M_i$ should be accessed to find the top-k final matches. So, the time complexity of the TA-based assembly in the worst case is $O(\sum_i(|M_i|))$. In Section \ref{effective}, we show the impact of the TA-based assembly on the performance.

\section{Approximate Optimization}
\label{bound}
In this section, we introduce an approximate optimization on the semantic-guided search to enable a trade-off between accuracy and the system response time (SRT) within a user-specified time bound $\mathcal{T}$. In the original A* semantic search, we will continuously check the priority queue $q$ until no partial pathes in $q$ having the greater estimated $pss$ than the explored matches' exact $pss$. This operation ensures that the output $k$ matches are globally optimal, but it also increases the SRT because the user cannot view the results before it terminates. Intuitively, if we can output the non-optimal matches earlier (within $\mathcal{T}$), then the SRT could be reduced. As more time is given, the non-optimal matches can be refined incrementally.

\begin{figure}
\setlength{\abovecaptionskip}{0cm}
\setlength{\belowcaptionskip}{-0.08cm}
\centerline{\includegraphics[scale=0.5]{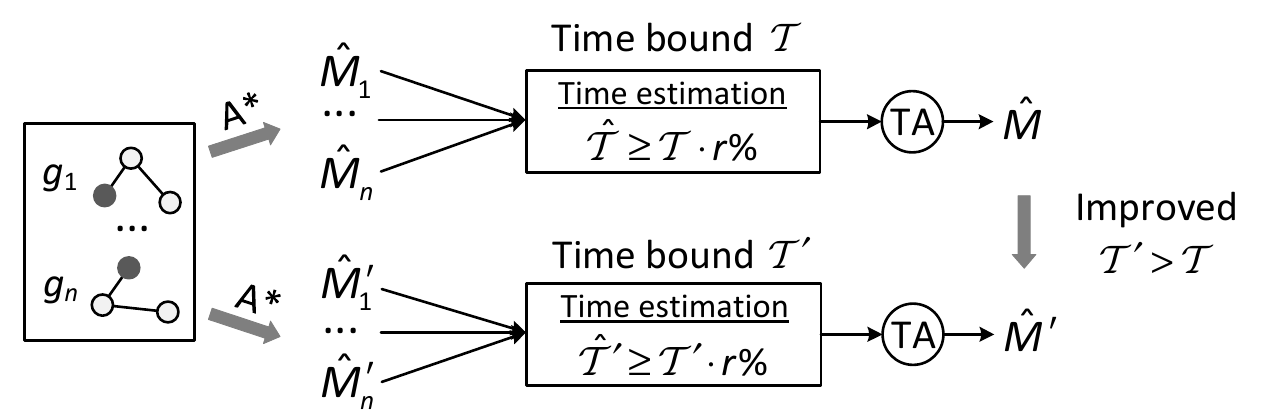}}
\caption{An example of approximate optimization}
\label{fig:ta1}
\end{figure}

\vspace{0.1cm}
\noindent\underline{\textbf{Core idea of the approximate optimization}}. Given a user-specific time bound $\mathcal{T}$, the approximate optimization is illustrated in Figure \ref{fig:ta1}. (1) We collect the early explored non-optimal matches of each sub-query graph $g_i$ to generate a non-optimal match set $\hat{M}_i$. (2) We estimate the possible overall time of assembling $\{\hat{M}_i\}$ to form the approximate final match set $\hat{M}$, denoted by $\hat{\mathcal{T}}$. (3) We decide to assemble $\hat{M}$, if $\hat{\mathcal{T}}\geq \mathcal{T}\cdot r\%$ is reached. The value of $\mathcal{T}\cdot r\%$ is an alert time threshold to indicate a level beyond which there is a risk of failing to return $\hat{M}$ within $\mathcal{T}$. Moreover, we ensure that $\hat{M}$ can be incrementally improved when more time is given. Two key differences between this optimization and the original A* semantic search (Algorithm \ref{alg:Asearch}) are provided as follows.

\begin{itemize}[leftmargin=*]
\item We collect each early explored match to the non-optimal match set $\hat{M}_i$ in the step of \textit{search space expansion} (lines 11-12 in Algorithm \ref{alg:Asearch1}).
\item We change the termination condition from \textit{top-k matches check} to \textit{execution time check} (lines 13-15 in Algorithm \ref{alg:Asearch1}). Specifically, we add a time estimation for the whole graph querying (Algorithm \ref{alg:Asearch1.1}) to ensure that the approximate final match set $\hat{M}$ can be returned within $\mathcal{T}$.
\end{itemize}

\noindent\underline{\textbf{Non-optimal match set $\hat{M}_i$}}. In Algorithm \ref{alg:Asearch1}, we update $\hat{M}_i$ once a match is explored no matter if it is optimal. Obviously, $\hat{M}_i$ will be incrementally refined as more time is given.

\begin{myLemma}
\label{lemma:optimal}
The non-optimal match set $\hat{M}_i$ can be incrementally refined and eventually be equal to the globally optimal match set $M_i$, if sufficient time is given.
\end{myLemma}

\begin{IEEEproof}
Suppose that we have $\hat{M}_i\cap M_i=\emptyset$ at time $t$, which means at least $|M_i|$ better matches are not explored in $\hat{M}_i$ so far. Hence, we have at least $|M_i|$ partial paths in the priority queue $q$ that have the greater estimated $pss$ $\hat{\psi}$ than the $pss$ of all matches in $\hat{M}_i$. According to Algorithm \ref{alg:Asearch1}, we will select the one with the greatest $\hat{\psi}$ from $q$ to expand the search space if $\hat{\mathcal{T}}<\mathcal{T}\cdot r\%$ holds. Once a match that belongs to $M_i$ is explored, we will collect it to $\hat{M}_i$, then we have $|\hat{M}_i\cap M_i|=1$. Theoretically, if sufficient time $\mathcal{T}$ is given, then Algorithm \ref{alg:Asearch1} will keep running until the best $M_i$ matches are explored.
\end{IEEEproof}


\begin{algorithm}[t]
\setstretch{0.8}
\small
\fontsize{7.8pt}{3.5mm}\selectfont
\caption{Time bounded A* semantic search}
\label{alg:Asearch1}
\KwData{sub-query graph $g_i$}
\KwResult{non-optimal match set $\hat{M}_i$}
$\forall u^s\in \phi(v^s)$:  $q$=$\{\langle u^s,\hat{\psi}_s\rangle\}$, \textit{visited}=$\{u_s\}$, $M_i$=$\emptyset$\;
\While{$q\neq \emptyset$}{
	$\langle\overline{u^su_i},\hat{\psi}_i\rangle$=$q$.pop\_max() \tcp*{\footnotesize{Next-hop selection}}
	\For(\tcp*[f]{\footnotesize{Search space expansion}}){$\forall u_l\in N(u_j)$}{
		\If{$!$visited.contains$(u_l)$}{
            \textit{visited}.add($u_l$)\;
			$\overline{u^su_l}$=$\overline{u^su_j}$+$u_ju_l$\;
			$\langle\overline{u^su_l},\hat{\psi}_l\rangle$=pssEstimation()\;
			\If{$\hat{\psi}_l\geq \tau$ and $u_l\not\in\phi(v^t)$}{
				$q$.push\_heap($\langle\overline{u^su_l},\hat{\psi}_l\rangle$)\;
			}
			\If{$\hat{\psi}_l\geq \tau$ and $u_l\in\phi(v^t)$}{
				$\hat{M}_i$.push\_heap($\langle\overline{u^su_l},\hat{\psi}_l\rangle$)\;
			}
		}
	}
	update($\mathcal{T}_{A^*}$)\;
	\If(\tcp*[f]{\footnotesize{Execution time check}}){timeEstimate$(\mathcal{T}_{A^*}$,$|\hat{M}_i|)$}{
		break\;
	}
}
\Return $\hat{M}_i$\;
\end{algorithm}
\setlength{\floatsep}{0cm}
\begin{algorithm}[t]
\setstretch{0.8}
\small
\fontsize{7.8pt}{3.5mm}\selectfont
\caption{timeEstimate($\mathcal{T}_{A^*}$,$|\hat{M}_i|$)}
\label{alg:Asearch1.1}
\KwData{$\langle\mathcal{T}_{A^*}$,$|\hat{M}_i|\rangle$ pair from each $g_i$}
collect the pair of $\langle\mathcal{T}_{A^*}$,$|\hat{M}_i|\rangle$ from each $g_i$\;
$\hat{\mathcal{T}}_{TA}$=$\sum{|\hat{M}_i|\cdot t}$, $\hat{\mathcal{T}}=max\{\mathcal{T}_{A^*}\}+\hat{\mathcal{T}}_{TA}$\;
\If{$\hat{\mathcal{T}}\geq\mathcal{T}\cdot r\%$}{
	\Return true\;
}
\Return false\;
\end{algorithm}

\vspace{0.05cm}
\noindent\underline{\textbf{Execution time check}}. The overall time for querying a query graph $G_Q$ is dominated by the time of A* semantic search ($\mathcal{T}_{A^*}$) for each sub-query graph $g_i$ and the time of TA-based assembly ($\mathcal{T}_{TA}$). Each $g_i$ is processed as an independent thread, so we use $max\{\mathcal{T}_{A^*}\}$ to denote the time of A* semantic search. Given a user-specific time bound $\mathcal{T}$, we want to obtain an approximate final match set $\hat{M}$ with the time $max\{\mathcal{T}_{A^*}\}+\mathcal{T}_{TA}\leq \mathcal{T}$. To this end, we estimate the overall time $\hat{\mathcal{T}}$ of our approximate optimization in Algorithm \ref{alg:Asearch1.1}. (1) The A* semantic search of each sub-query graph $g_i$ reports a pair of $\langle\mathcal{T}_{A^*}$,$|\hat{M}_i|\rangle$ for time estimation (line 1), where $\mathcal{T}_{A^*}$ is the current running time of each $g_i$ and $|\hat{M}_i|$ is the number of explored matches so far. (2) We use all the $|\hat{M}_i|$ to estimate $\mathcal{T}_{TA}$ (line 2). In the worst case, TA-based assembly needs to access all matches in $\hat{M}_i$, so we use $\sum{|\hat{M}_i|\cdot t}$ as the estimated $\hat{\mathcal{T}}_{TA}$, where $t$ is an empirical time for processing one match of $\hat{M}_i$ in the TA-based assembly. (3) We decide to launch the TA-based assembly if $max\{\mathcal{T}_{A^*}\}+\hat{\mathcal{T}}_{TA}\geq \mathcal{T}\cdot r\%$ (line 3). Otherwise, the A* semantic search will continue to update $\hat{M}_i$.


\vspace{0.05cm}
\noindent\underline{\textbf{Approximate $\hat{M}$ assembly}}. Given a set of non-optimal match sets $\{\hat{M}_i\}$, we conduct a TA-based assembly to generate the approximate final match set $\hat{M}$. In this paper, we use the Jaccard similarity between $\hat{M}$ and the globally optimal final match set $M$ to quantify their approximation degree as follows,

\vspace{-0.35cm}
\begin{equation}
\label{eq:jaccard}
Jad(\hat{M},M)=\frac{|\hat{M}\cap M|}{|\hat{M}\cup M|}=\frac{k_{\cap}}{2k-k_{\cap}}
\end{equation}
\vspace{-0.35cm}

\noindent where $k$ is the size of approximate final match set, and $k_{\cap}$ is the size of $|\hat{M}\cap M|$.


\begin{myTheorem}
\label{th:jaccard}
Our approximate optimization can incrementally refine the approximate final match set $\hat{M}$ and finally obtain the globally optimal $M$, if sufficient time is given.
\end{myTheorem}

\begin{IEEEproof}
Suppose that we have a non-optimal match set $\hat{M}_i$ at time $t$. According to Lemma \ref{lemma:optimal}, $\hat{M}_i$ can be refined to $\hat{M}_i'$ at time $t'$ ($t'>t$). Hence, the approximate final match set $\hat{M}'$ assembled from $\{\hat{M}_i'\}$ is better than $\hat{M}$ assembled from $\{\hat{M}_i\}$, which means $k_{\cap}'\geq k_{\cap}$. According to Eq. \ref{eq:jaccard}, $Jad(\hat{M}',M)\geq Jad(\hat{M},M)$ holds if $k_{\cap}'\geq k_{\cap}$. Moreover, if sufficient time $\mathcal{T}$ is given, then we have $\hat{M}_i=M_i$ (Lemma \ref{lemma:optimal}). Hence, the global optimal final match set $M$ can be assembled from $\{\hat{M}_i\}$ when sufficient time $\mathcal{T}$ is given.
\end{IEEEproof}

\begin{table}
\setstretch{0.8}
\fontsize{6.8pt}{2.8mm}\selectfont
\setlength{\abovecaptionskip}{0cm}
  \centering
  \caption{Statistics of datasets}
  \begin{tabular} {c||c|c|c|c}
   \hline
   Datasets & \# Nodes & \# Edges & \# Node-Types & \# Edge-Predicates\\
   \hline \hline
   DBpedia & 4,521,912 & 15,045,801 & 359 & 676\\
   \hline
   Freebase & 5,706,539 & 48,724,743 & 11,666 & 5118\\
   \hline
   YAGO2 & 7,308,372 & 36,624,106 & 6,543 & 101\\
   \hline
  \end{tabular}
  \label{tab:dataset}
\end{table}

\section{Experimental Study}
\label{exp}
We present experiment results to evaluate (1) effectiveness and efficiency, (2) analysis via user-study, (3) impact of query graph shapes and sizes, (4) robustness with noise, (5) scalability of our algorithms, and (6) parameter sensitivity. The source code of this paper can be obtained from {\footnotesize https://github.com/hqf1996/Semantic-guided-search}.

\subsection{Experimental Setup}
\label{setup}
\noindent\underline{\textbf{Datasets.}} We used three real-world datasets as shown in Table \ref{tab:dataset}. (1) \textbf{\textit{DBpedia}} \cite{Mendes2012} is an open-domain knowledge base, which is constructed from Wikipedia. We used the same DBpedia dataset as \cite{Zheng2016} (authors shared it with us). (2) \textbf{\textit{Freebase}} \cite{Bollacker2007} is a knowledge base mainly composed by communities. Since we assume that each entity has a name, we used a Freebase-Wikipedia mapping file \cite{freebaselinks} to filter 5.7M entities, each entity has a name from Wikipedia. (3) \textbf{\textit{YAGO2}} \cite{Hoffart2013} is a knowledge base with information from the
Wikipedia, WordNet and GeoNames. In this paper, we only used the {\sf CORE} portion of Yago (excluding information from GeoName) as our dataset.

\begin{table*}
\setlength{\abovecaptionskip}{0.05cm}
\setstretch{0.8}
\fontsize{6.8pt}{3mm}\selectfont
    \centering
    \tabcaption{Effectiveness and efficiency over DBpedia (top-$k=20, 40, 100, 200$)}
    \begin{tabular}{c||c|c|c|c|c|c|c|c|c|c|c|c|c|c|c|c}
    \hline
    \multirow{2}*{\textbf{Methods}} & \multicolumn{4}{c|}{\textbf{Precision}} & \multicolumn{4}{c|}{\textbf{Recall}} & \multicolumn{4}{c|}{\textbf{F1-measure}} & \multicolumn{4}{c}{\textbf{Query Answering Time (ms)}}\\
    \cline{2-17}
    & $k$=20 & $k$=40 & $k$=100 & $k$=200 & $k$=20 & $k$=40 & $k$=100 & $k$=200 & $k$=20 & $k$=40 & $k$=100 & $k$=200 & $k$=20 & $k$=40 & $k$=100 & $k$=200\\
    \hline \hline
    SGQ & \textbf{0.94} & \textbf{0.96} & \textbf{0.96} & \textbf{0.88} & \textbf{0.09} & \textbf{0.19} & \textbf{0.48} & \textbf{0.69} & \textbf{0.17} & \textbf{0.31} & \textbf{0.59} & \textbf{0.73} & \textbf{65.41} & \textbf{73.26} & \textbf{102.20} & \textbf{136.81}\\
    \hline
    TBQ-0.9 & 0.87 & 0.92 & 0.91 & 0.83 & 0.08 & 0.18 & 0.45 & 0.67 & 0.12 & 0.28 & 0.57 & 0.70 & \textbf{54.23} & \textbf{69.17} & \textbf{93.86} & \textbf{122.69}\\
    \hline
    S4 & 0.61 & 0.70 & 0.81 & 0.76 & 0.06 & 0.15 & 0.40 & 0.61 & 0.11 & 0.24 & 0.49 & 0.65 & 185.80 & 224.61 & 352.35 & 385.22\\
    \hline
    GraB & 0.79 & 0.82 & 0.85 & 0.71 & 0.08 & 0.17 & 0.44 & 0.57 & 0.15 & 0.27 & 0.54 & 0.59 & 382.69 & 386.07 & 470.47 & 651.76\\
    \hline
    QGA & 0.79 & 0.65 & 0.47 & 0.33 & 0.08 & 0.13 & 0.24& 0.36 & 0.15 & 0.22 & 0.32 & 0.34 & 787.80 & 821.93 & 1127.41 & 1303.35\\
    \hline
    $p$-hom & 0.37 & 0.32 & 0.29 & 0.27 & 0.05 & 0.08 & 0.18 & 0.33 & 0.10 & 0.13 & 0.22 & 0.30 & 1214.33 & 1216.83 & 1243.5 & 1269.67\\
    \hline
    \end{tabular}
    \vspace{1mm}
    \label{tab:effectiveness1}
\end{table*}

\begin{table*}
\setlength{\abovecaptionskip}{0.05cm}
\setlength{\belowcaptionskip}{-0.3cm}
\setstretch{0.8}
\fontsize{6.8pt}{3mm}\selectfont
    \centering
    \tabcaption{Effectiveness and efficiency over Freebase (top-$k=20, 40, 100, 200$)}
    \begin{tabular}{c||c|c|c|c|c|c|c|c|c|c|c|c|c|c|c|c}
    \hline
    \multirow{2}*{\textbf{Methods}} & \multicolumn{4}{c|}{\textbf{Precision}} & \multicolumn{4}{c|}{\textbf{Recall}} & \multicolumn{4}{c|}{\textbf{F1-measure}} & \multicolumn{4}{c}{\textbf{Query Answering Time (ms)}}\\
    \cline{2-17}
    & $k$=20 & $k$=40 & $k$=100 & $k$=200 & $k$=20 & $k$=40 & $k$=100 & $k$=200 & $k$=20 & $k$=40 & $k$=100 & $k$=200 & $k$=20 & $k$=40 & $k$=100 & $k$=200\\
    \hline \hline
    SGQ & \textbf{0.95} & \textbf{0.95} & \textbf{0.87} & \textbf{0.73} & \textbf{0.12} & \textbf{0.24} & \textbf{0.52} & \textbf{0.74} & \textbf{0.20} & \textbf{0.36} & \textbf{0.62} & \textbf{0.68} & \textbf{115.20} & \textbf{118.82} & \textbf{147.30} & \textbf{212.83}\\
    \hline
    TBQ-0.9 & 0.91 & 0.90 & 0.83 & 0.68 & 0.11 & 0.21 & 0.49 & 0.68 & 0.19 & 0.31 & 0.60 & 0.62 & \textbf{100.57} & \textbf{113.87} & \textbf{133.03} & \textbf{191.05}\\
    \hline
    S4 & 0.79 & 0.76 & 0.65 & 0.64 & 0.09 & 0.14 & 0.38 & 0.62 & 0.14 & 0.23 & 0.45 & 0.55 & 185.82 & 212.60 & 271.43 & 342.28\\
    \hline
    GraB & 0.85 & 0.87 & 0.63 & 0.54 & 0.11 & 0.18 & 0.36 & 0.55 & 0.20 & 0.29 & 0.43 & 0.50 & 345.60 & 324.80 & 357.82 & 564.75\\
    \hline
    QGA & 0.88 & 0.82 & 0.60 & 0.46 & 0.11 & 0.16 & 0.33 & 0.42 & 0.20 & 0.26 & 0.43 & 0.44 & 811.98 & 1029.66 & 1567.81 & 1606.24\\
    \hline
    $p$-hom & 0.35 & 0.26 & 0.23 & 0.22 & 0.06 & 0.09 & 0.14 & 0.26 & 0.10 & 0.13 & 0.17 & 0.24 & 996.22 & 1016.22 & 1077.34 & 1125.78\\
    \hline
    \end{tabular}
    \vspace{1mm}
    \label{tab:effectiveness2}
\end{table*}

\begin{table*}
\setlength{\abovecaptionskip}{0.05cm}
\setlength{\belowcaptionskip}{-0.3cm}
\setstretch{0.8}
\fontsize{6.8pt}{3mm}\selectfont
    \centering
    \tabcaption{Effectiveness and efficiency over YAGO2 (top-$k=20, 40, 100, 200$)}
    \begin{tabular}{c||c|c|c|c|c|c|c|c|c|c|c|c|c|c|c|c}
    \hline
    \multirow{2}*{\textbf{Methods}} & \multicolumn{4}{c|}{\textbf{Precision}} & \multicolumn{4}{c|}{\textbf{Recall}} & \multicolumn{4}{c|}{\textbf{F1-measure}} & \multicolumn{4}{c}{\textbf{Query Answering Time (ms)}}\\
    \cline{2-17}
    & $k$=20 & $k$=40 & $k$=100 & $k$=200 & $k$=20 & $k$=40 & $k$=100 & $k$=200 & $k$=20 & $k$=40 & $k$=100 & $k$=200 & $k$=20 & $k$=40 & $k$=100 & $k$=200\\
    \hline \hline
    SGQ & \textbf{0.75} & \textbf{0.76} & \textbf{0.73} & \textbf{0.69} & \textbf{0.04} & \textbf{0.07} & \textbf{0.17} & \textbf{0.33} & \textbf{0.07} & \textbf{0.13} & \textbf{0.28} & \textbf{0.45} & \textbf{113.41} & \textbf{118.80} & \textbf{131.04} & \textbf{147.42}\\
    \hline
    TBQ-0.9 & 0.72 & 0.74 & 0.71 & 0.67 & 0.03 & 0.07 & 0.17 & 0.32 & 0.07 & 0.13 & 0.27 & 0.43 & \textbf{102.06} & \textbf{112.46} & \textbf{124.80} & \textbf{143.20}\\
    \hline
    S4 & 0.64 & 0.67 & 0.65 & 0.63 & 0.03 & 0.06 & 0.15 & 0.30 & 0.06 & 0.12 & 0.25 & 0.41 & 190.64 & 212.73 & 275.11 & 364.38\\
    \hline
    GraB & 0.60 & 0.64 & 0.62 & 0.59 & 0.03 & 0.06 & 0.15 & 0.28 & 0.05 & 0.11 & 0.24 & 0.38 & 287.17 & 302.25 & 468.01 & 495.33\\
    \hline
    QGA & 0.65 & 0.60 & 0.57 & 0.57 & 0.03 & 0.06 & 0.14 & 0.28 & 0.06 & 0.11 & 0.23 & 0.37 & 928.84 & 949.32 & 982.68 & 1017.32\\
    \hline
    $p$-hom & 0.35 & 0.40 & 0.36 & 0.34 & 0.02 & 0.04 & 0.09 & 0.17 & 0.03 & 0.07 & 0.15 & 0.23 & 622.50 & 692.45 & 717.51 & 1065.23\\
    \hline
    \end{tabular}
    \vspace{1mm}
    \label{tab:effectiveness3}
\end{table*}

\vspace{0.05cm}
\noindent\underline{\textbf{Query workload.}} We used four query workloads to construct the query graphs. (1) \textbf{\textit{QALD-4}} \cite{qald} is a benchmark for \textit{DBpedia}. It provides both SPARQL expression and answers for each query. A SPARQL expression may involve multiple UNION operators, which correspond to different predefined schemas in DBpedia. We selected only one UNION operator to construct the query graph. It is desired to find more answers without considering all UNION operators. (2) \textbf{\textit{WebQuestions}} \cite{Berant2013} is a benchmark for \textit{Freebase}. It provides a set of questions, denoted by a quadruple $\langle$\textit{qText}, \textit{freebaseKey}, \textit{relPaths}, \textit{answers}$\rangle$. We took the entities and relations from \textit{freebaseKey} and \textit{relPaths} to form query graphs. (3) \textbf{\textit{RDF-3x}} \cite{Neumann2008} contains queries for \textit{YAGO} dataset. It provides SPARQL expressions, but does not provide the answers. To obtain the validation set, we imported YAGO2 to the graph database Neo4j and executed the queries through the sparql-plugin. (4) \textbf{\textit{Synthetic graphs}} were generated to evaluate the effect of query graph shapes and sizes on our approach. According to \cite{Bonifati2017}, chain, star, tree, cycle, and flower are the commonly used graph shapes, so we generated query graphs with these shapes by extracting subgraphs from our datasets. Similar to \cite{Bonifati2017}, we took the number of edges (i.e., triples) in the query graph to measure the query size.


\vspace{0.05cm}
\noindent\underline{\textbf{Metrics.}} We adopted two classical metrics to measure the effectiveness. \textit{Precision} ($P$) is the ratio of  correctly discovered answers over all discovered top-k answers. \textit{Recall} ($R$) is the ratio of  correctly discovered answers over all correct answers. In addition, we also employed \textit{F1-measure} to combine the precision and recall as $F1=\frac{2}{1/P+1/R}$.

\vspace{0.05cm}
\noindent\underline{\textbf{Comparing methods.}} We compared our approach with four recent works on knowledge graph search: S4 \cite{Zheng2016}, $p$-hom \cite{Fan2010a}, GraB \cite{Jin2015}, and QGA \cite{Han2017}. S4 is a semantic pattern based solution. $p$-hom and GraB support structurally edge-to-path mapping. QGA is a keyword based approach.



There are two versions of our approach: (1) SGQ (semantic-guided query) is the implementation of A* semantic search and TA assembly in Section \ref{Asearch}. (2) TBQ (time-bounded query) is the approximate optimization in Section \ref{bound}. We set the default $pss$ threshold $\tau$=0.8 and user desired path length $\hat{n}$=4. We used the TransE \cite{Bordes2013}  embedding model to obtain the predicate semantic space. All the experiments were conducted on a 2.1GHZ, 64GB memory AMD-6272 server with a single core.

\begin{table}
\setlength{\abovecaptionskip}{0.05cm}
\setlength{\belowcaptionskip}{-0.5cm}
 \fontsize{6.8pt}{2.8mm}\selectfont
  \centering
  \caption{The schemas of returned answers for Q117}
  \begin{tabular}{l}
   \hline
   \textbf{Answers' schemas of $G_Q$:} {\begin{minipage}{0.28\textwidth}
      \centerline{\includegraphics[scale=0.46]{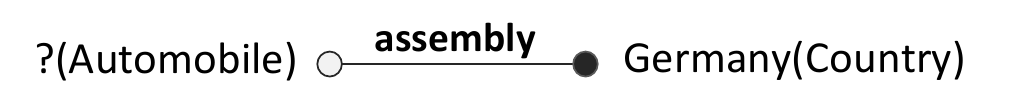}}
    \end{minipage}}\\
   \hline \hline
   \textit{Automobile}--\textbf{assembly}--\textit{Germany}\\
   \textit{Automobile}--\textbf{assembly}--\textit{City}--\textbf{country}--\textit{Germany}\\
   \textit{Automobile}--\textbf{manufacturer}--\textit{Company}--\textbf{location}--\textit{Germany}\\
   \textit{Automobile}--\textbf{manufacturer}--\textit{Company}--\textbf{locationCountry}--\textit{Germany}\\
   \hline
   \textit{Automobile}--\textbf{assembly}--\textit{Company}--\textbf{location}--\textit{Germany}\\
   \textit{Automobile}--\textbf{assembly}--\textit{Company}--\textbf{locationCountry}--\textit{Germany}\\
   \textit{Automobile}--\textbf{designCompany}--\textit{Company}--\textbf{location}--\textit{Germany}\\
   \hline
  \end{tabular}
  \vspace{1mm}
  \label{tab:example}
\end{table}

\vspace{-0.15cm}
\subsection{Effectiveness and Efficiency Evaluation}
\label{effective}
\vspace{-0.05cm}
\noindent\underline{\textbf{Effectiveness.}} In this test, we set the time bound of TBQ as 90\% of the execution time of SGQ (TBQ-0.9). Tables \ref{tab:effectiveness1}-\ref{tab:effectiveness3} (\textit{Precision}, \textit{Recall}, and \textit{F1-measure}) show the effectiveness results over different top-k. For all datasets, our approach outperforms the others. This is because we can find the semantically similar answers following the guidance of the predicate semantics. Table \ref{tab:example} shows some schemas of returned answers for the example query in Figure \ref{fig:example2} (Q117 from QALD-4). Our approach can find the correct answers (e.g., with the first four schemas). It also finds some reasonable answers not given in the validation set (e.g., with the last three schemas).

\vspace{0.05cm}
\noindent\underline{\textbf{Efficiency.}} Tables \ref{tab:effectiveness1}-\ref{tab:effectiveness3} (\textit{Time}) report that our approach outperforms the other methods because unpromising answers are pruned significantly through the effective $pss$ estimation in runtime. It is natural that delivering more answers (larger $k$) consumes more search time. Moreover, we provide the average time of each component of our approach in Table \ref{tab:component}: \textit{query graph decomposition} (C1), \textit{semantic graph construction} (C2), \textit{semantic-guided search} (C3), and \textit{TA-based assembly} (C4). For C2, we show the time of constructing semantic graph partially and completely. While in C3, we show the time of A* semantic search with or without $pss$ estimation. According to the results (bold), we observe that our solutions adopted in C2 and C3 outperform the baselines, which proves that the partially semantic graph construction and $pss$ estimation are very helpful to reduce the search space and improve the efficiency. Among four components, the most time-consuming one is C3. This is because we need to estimate $pss$ for each candidate node match explored in the edge-to-path mapping.




\begin{table}
\setlength{\abovecaptionskip}{0.05cm}
\setlength{\belowcaptionskip}{-0.5cm}
\setstretch{0.8}
\fontsize{6.8pt}{2.8mm}\selectfont
  \centering
  \caption{Average time (ms) of each component (top-$k$=100). C1: Query graph decomposition, C2: Semantic graph construction, C3: Semantic-guided search, and C4: TA-based assembly.}
  \begin{tabular} {c||c|c|c|c|c|c}
   \hline
   \multirow{2}*{\textbf{Dataset}} & \multirow{2}{*}{C1} & \multicolumn{2}{c|}{C2} & \multicolumn{2}{c|}{C3} & \multirow{2}{*}{C4} \\
   \cline{3-6}
   & & partial & complete & prune & w/o prune & \\
   \hline \hline
   DBpedia & 4.30 & \textbf{6.59} & 157.33 & \textbf{88.11} & 435.67 & 3.20 \\
   \hline
   Freebase & 4.10 & \textbf{12.84} & 274.87 & \textbf{126.16} & 701.20 & 4.20 \\
   \hline
   Yago2 & 7.05 & \textbf{9.46} & 121.80 & \textbf{106.91} & 404.30 & 7.62 \\
   \hline
  \end{tabular}
  \vspace{1mm}
  \label{tab:component}
\end{table}
\setlength{\floatsep}{-0.2cm}
\begin{figure}
\setlength{\abovecaptionskip}{-0.1cm}
\setlength{\belowcaptionskip}{0cm}
\centering
\subfigcapskip=-0.2cm
\subfigure[Effectiveness for TBQ]{
\begin{minipage}[t]{0.46\linewidth}
\centering{\includegraphics[height=2.4cm, width=3.4cm]{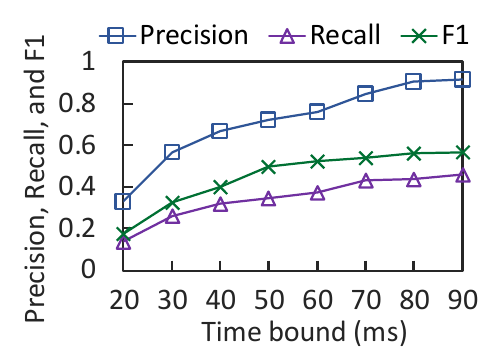}}
\label{fig:tbq1}
\end{minipage}
}
\subfigure[Efficiency for TBQ]{
\begin{minipage}[t]{0.46\linewidth}
\centering{\includegraphics[height=2.4cm, width=3.4cm]{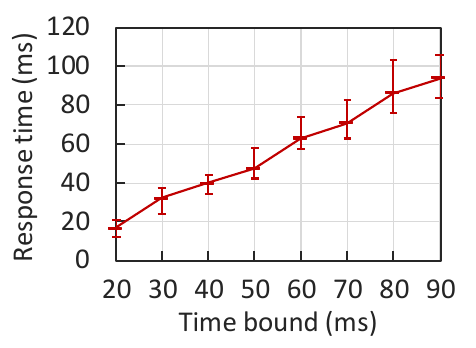}}
\label{fig:tbq2}
\end{minipage}
}
\caption{Impact of response time bounds (DBpedia, top-$k$=100)}
\label{fig:tbq}
\end{figure}

\vspace{0.05cm}
\noindent\underline{\textbf{Response Time-Accuracy Trade-off.}} Figure \ref{fig:tbq} reports the effect of time bounds on TBQ. Because the results over three datasets show the similar trends, we only provide the results over DBpedia for top-$k$=100. We varied the time bound from 20 ms to 90 ms to evaluate the effectiveness and efficiency of TBQ. Figure \ref{fig:tbq1} shows that more accurate answers can be returned as more time is given. In Figure \ref{fig:tbq2}, each bar represents the minimum, maximum, and average response times
of queries. Observe that, TBQ can return the answers within a small variation of the actual time bound provided.

\vspace{0.05cm}
\noindent\underline{\textbf{Optimal Query Graph Decomposition Benefit.}} Given the same query graph $G_Q$, different query graph decompositions may, however, cause different runtimes over the same dataset. We show the total runtime of our approach over three datasets based on the optimal query graph decomposition and the worst case decomposition in Table \ref{tab:benefit}. For instance, the runtime of the optimal query graph decomposition and the worst case decomposition over DBpedia are 102.2 ms and 122.74 ms, respectively. The efficiency is improved by 16.73\%. The results demonstrate the usefulness of our DP-based optimal query graph decomposition.

\begin{table}
\setlength{\abovecaptionskip}{0.05cm}
\setstretch{0.8}
\fontsize{6.8pt}{2.8mm}\selectfont
  \centering
  \caption{Total runtime (ms) of different decomposition cases}
  \begin{tabular} {c||c|c|c}
   \hline
   \textbf{Different cases}  & \textbf{DBpedia} & \textbf{Freebase} & \textbf{Yago2}\\
   \hline \hline
   optimal case & 102.20 & 147.30 & 131.04\\
   \hline
   worst case & 122.74 & 180.29 & 194.86\\
   \hline
  \end{tabular}
  \vspace{1mm}
  \label{tab:benefit}
\end{table}

\vspace{-0.2cm}
\subsection{User Study}
\label{userstudy}
\vspace{-0.1cm}
Since our approach returns the top-k answers to the user, we want to know if users are satisfied with the high-ranking answers (even though the answers are already in the validation set). We expect that an answer that is more familiar to the user must have a higher rank. Therefore, we conducted a user study through \textit{Baidu Data CrowdSourcing Platform} ({\footnotesize https://zhongbao.baidu.com/?language=en}) to evaluate the correlation between top-k answers of our approach (SGQ) and user's preference, measured by Pearson Correlation Coefficient (PCC). We did this test according to the steps in \cite{Jayaram2015}. We selected 20 queries (6, 12, and 2 queries from QALD-4, WebQuestions, and RDF-3x, respectively) for this user study. For each query, we generated 30 random pairs of answers (two answers from the same pair have different schemas), and presented each pair to 10 annotators and asked for their preference (Table \ref{tab:PCC1} for example). If more higher ranked answers (e.g., \textit{BMW\_320}) are preferred by more users, then we can say that the top-k answers and users' preferences are positively correlated.

\begin{table}
\setlength{\abovecaptionskip}{0.05cm}
\setlength{\belowcaptionskip}{0.2cm}
\setstretch{0.8}
\fontsize{6.8pt}{2.8mm}\selectfont
  \centering
  \caption{An example of answer pairs for user study (Q117)}
  \begin{tabular} {c|c|c||c|c|c}
   \hline
   \textbf{Answer1} & \textbf{SGQ rank} & \textbf{User} & \textbf{Answer2} & \textbf{SGQ rank} & \textbf{User}\\
   \hline \hline
   Opel\_Super\_6 & 62 &  & BMW\_320 & 10 & \tiny{\CheckmarkBold}\\
   \hline
   Volkswagen\_Passat & 72 & \tiny{\CheckmarkBold} & 30\_PS & 26 & \\
   \hline
  \end{tabular}
  \vspace{1mm}
  \label{tab:PCC1}
\end{table}

\begin{table}[!t]
\setlength{\abovecaptionskip}{0.05cm}
\setlength{\belowcaptionskip}{0.2cm}
\setstretch{0.8}
\fontsize{6.8pt}{2.8mm}\selectfont
  \centering
  \caption{PCC results (DBpedia (D), Freebase (F), YAGO2 (Y))}
  \begin{tabular} {c|c||c|c||c|c||c|c}
   \hline
   \textbf{Query} & \textbf{PCC} & \textbf{Query} & \textbf{PCC} & \textbf{Query} & \textbf{PCC} & \textbf{Query} & \textbf{PCC}\\
   \hline \hline
   D1 & 0.46 & D6 & 0.74 & F5 & 0.69 & F10 & 0.73\\
   \hline
   D2 & 0.56 & F1 & 0.74 & F6 & 0.37 & F11 & 0.69\\
   \hline
   D3 & 0.61 & F2 & 0.72 & F7 & 0.41 & F12 & 0.77\\
   \hline
   D4 & 0.75 & F3 & 0.77 & F8 & 0.71 & Y1 & 0.74\\
   \hline
   D5 & 0.73 & F4 & 0.72 & F9 & 0.74 & Y2 & 0.45\\
   \hline
  \end{tabular}
  \vspace{1mm}
  \label{tab:PCC}
\end{table}

Finally, we obtained 20*30*10=6000 opinions in total. We constructed two lists $X$ and $Y$ for each query based on these opinions. Each list has 30 values for 30 answer pairs. For each pair, the value in $X$ is the difference between the two answers' ranks given by SGQ, and the value in $Y$ is the difference between the numbers of annotators favoring the two answers. Then, we calculated the PCC for each query based on Eq. \ref{eq:pcc}. The PCC value shows the degree of correlation between the preference given by SGQ and annotators. A PCC value in the ranges of [0.5,1.0], [0.3,0.5) and [0.1,0.3) indicates a strong, medium and small positive correlation, respectively \cite{Jayaram2015}. Table \ref{tab:PCC} shows that SGQ achieved strong and medium positive correlations with the annotators on 16 and 4 queries, respectively, which indicates that the users were satisfied with the semantically similar answers identified via our approach.


\vspace{-0.2cm}
\begin{equation}
\fontsize{9pt}{2.8mm}\selectfont
\label{eq:pcc}
PCC=\frac{E(XY)-E(X)\cdot E(Y)}{\sqrt{E(X^2)-E(X)^2}\cdot \sqrt{E(Y^2)-E(Y)^2}}
\end{equation}
\vspace{-0.4cm}

\vspace{-0.1cm}
\subsection{Effect of Query Graph Shapes and Sizes}
\label{pivot}
\vspace{-0.1cm}
In this experiment, we evaluated the effect of query graph shapes and sizes. (1) \textbf{\textit{Graph shapes}}. We extracted the subgraphs from the original knowledge graph as the query graphs with different shapes, e.g., chain, star, tree, cycle, and flower (they are very common in knowledge graph search \cite{Bonifati2017}). Figure \ref{fig:shape} shows a \textit{flower}-shaped query graph and its definition provided in \cite{Bonifati2017}. (2) \textbf{\textit{Graph sizes}}. Similar to \cite{Bonifati2017}, we use the number of edges (i.e., triples) in a query graph to indicate the size of a query graph, including \textit{\textbf{S}mall} ($1\leq |E_Q|\leq 4$) and \textit{\textbf{L}arge} ($5\leq |E_Q|\leq 10$). (3) \textbf{\textit{Specific nodes}}. For each query graph, we randomly select 2 to 4 query nodes as specific nodes and others are target nodes. For each pair of $\langle$shape,$|E_Q|\rangle$, we have 5 query graphs. Table \ref{tab:shape} shows the experimental results.

\vspace{0.1cm}
\noindent\underline{\textbf{Effect of shapes}}. The tree and flower shaped query graphs are more time-consuming than others especially for the large query sizes (e.g., 1467 ms on average for large flower graphs). This is because they have more sub-query graphs than other cases. For example, the larger flower-shaped query graphs have the most sub-query graphs (4.62 on average).

\vspace{0.05cm}
\noindent\underline{\textbf{Effect of sizes}}. The large query graphs always take more time than small ones for all shapes. Since the large query graphs have more edges, more candidate node matches need to be considered in the edge-to-path mapping, which increases the time of \textit{semantic graph construction} (C2) and \textit{semantic-guided search} (C3). For \textit{query graph decomposition} (C1), its running time is much smaller compared with C2 and C3. For instance, C1 takes only 12.47 ms on average for the large flower-shaped query graphs. Even for a quite large flower-shaped query (with 10 edges), it takes only 18.79 ms. Moreover, 90.76\% of the query graphs have at most 6 edges \cite{Bonifati2017}, so we can say that C1 is scalable to the query size in practice. While for \textit{TA-based assembly} (C4), the time complexity in the worst case is $O(\sum_i|M_i|)$, which is dominated by the number of sub-query graphs and $|M_i|$. If we want to find the top-k matches (e.g., $k$=100), the $|M_i|$ is usually the same order of magnitude as $k$. Besides, the number of sub-query graphs is usually small in practice (e.g., 4.62 on average for the large flower-shaped query graphs). So, the size of $\sum_i|M_i|$ is not large in practice. Hence, we conclude that C4 is also scalable to the query size.


\begin{figure}
\setlength{\abovecaptionskip}{-0.1cm}
\setlength{\belowcaptionskip}{0cm}
\centering
\subfigcapskip=-0.2cm
\subfigure[Node noise]{
\begin{minipage}{0.45\linewidth}
\centering{\includegraphics[scale=0.78]{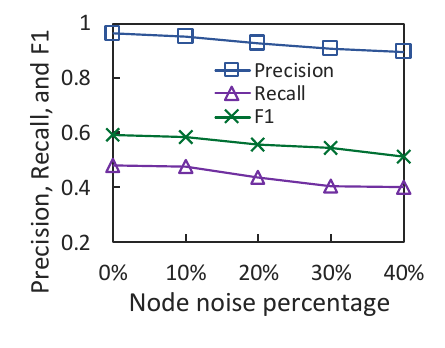}}
\label{fig:noise1}
\end{minipage}
}
\subfigure[Edge noise]{
\begin{minipage}{0.45\linewidth}
\centering{\includegraphics[scale=0.78]{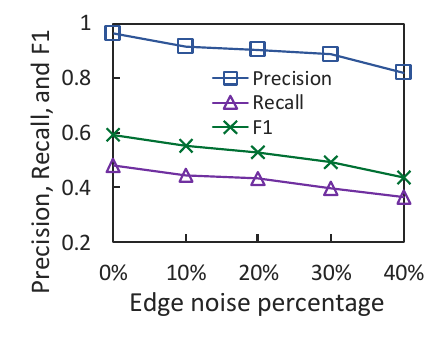}}
\label{fig:noise2}
\end{minipage}
}
\caption{Effectiveness vs. Noise (DBpedia, top-$k$=100)}
\label{fig:noise}
\end{figure}

\begin{table}
\setlength{\abovecaptionskip}{0.05cm}
\setlength{\belowcaptionskip}{0.2cm}
\setstretch{0.8}
\fontsize{6.8pt}{2.8mm}\selectfont
  \centering
  \caption{Response time (ms) vs. Noise (DBpedia, top-$k$=100)}
  \begin{tabular} {c||c|c|c|c|c}
   \hline
   \textbf{Noise type} & 0\% & 10\% & 20\% & 30\% & 40\%\\
   \hline \hline
   node noise & 102.2 & 105.3 & 114.6 & 117.3 & 126.7 \\
   \hline
   edge noise & 102.2 & 116.5 & 126.7 & 145.1 & 168.6 \\
   \hline
  \end{tabular}
  \vspace{1mm}
  \label{tab:noise}
\end{table}

\begin{figure*}
\setlength{\abovecaptionskip}{0.05cm}
\setstretch{0.8}
\fontsize{6pt}{2.8mm}\selectfont
  \begin{minipage}[b]{0.65\textwidth}
    \centering
    \tabcaption{Analysis of query graph shapes and complexity (DBpedia, Response time (ms))}
    \begin{tabular}
    {p{1.6cm}<{\centering}||p{0.48cm}<{\centering}|p{0.55cm}<{\centering}|p{0.5cm}<{\centering}|p{0.55cm}<{\centering}|p{0.5cm}<{\centering}|p{0.55cm}<{\centering}|p{0.5cm}<{\centering}|p{0.55cm}<{\centering}|p{0.5cm}<{\centering}|p{0.55cm}<{\centering}}
    \hline
    \multirow{2}*{\textbf{Metric}} & \multicolumn{2}{c|}{Chain} & \multicolumn{2}{c|}{Cycle} & \multicolumn{2}{c|}{Star} & \multicolumn{2}{c|}{Tree} & \multicolumn{2}{c}{Flower} \\
    \cline{2-11}
    & S & L & S & L & S & L & S & L & S & L \\
    \hline \hline
    Time (C1) & 4.30 & 7.57 & 7.40 & 11.00 & 7.20 & 9.10 & 6.70 & 9.87 & 9.00 & 12.47\\
    \hline
    Time (C2) & 8.80 & 24.97 & 19.80 & 35.50 & 20.50 & 54.37 & 24.20 & 87.47 & 38.20 & 132.36\\
    \hline
    Time (C3) & 111.70 & 687.40 & 229.00 & 762.03 & 280.30 & 834.50 & 255.90 & 1116.73 & 359.00 & 1314.90\\
    \hline
    Time (C4) & 2.80 & 6.40 & 7.90 & 5.47 & 2.90 & 5.90 & 3.40 & 7.77 & 7.40 & 8.10\\
    \hline
    Total & 127.60 & 726.33 & 264.10 & 814.00 & 310.90 & 903.87 & 290.20 & 1221.83 & 413.60 & 1467.83\\
    \hline
    Precision & 0.89 & 0.81 & 0.90 & 0.83 & 0.89 & 0.84 & 0.92 & 0.85 & 0.84 & 0.82\\
    \hline
    \#sub-query (avg.) & 1.25 & 2.00 & 2.00 & 2.00 & 3.12 & 3.50 & 3.00 & 3.83 & 3.00 & 4.62\\
    \hline
    \end{tabular}
    \vspace{1mm}
    \label{tab:shape}
  \end{minipage}
  \begin{minipage}[b]{0.35\textwidth}
  \setlength{\abovecaptionskip}{0.3cm}
    \centering
    \includegraphics[scale=0.5]{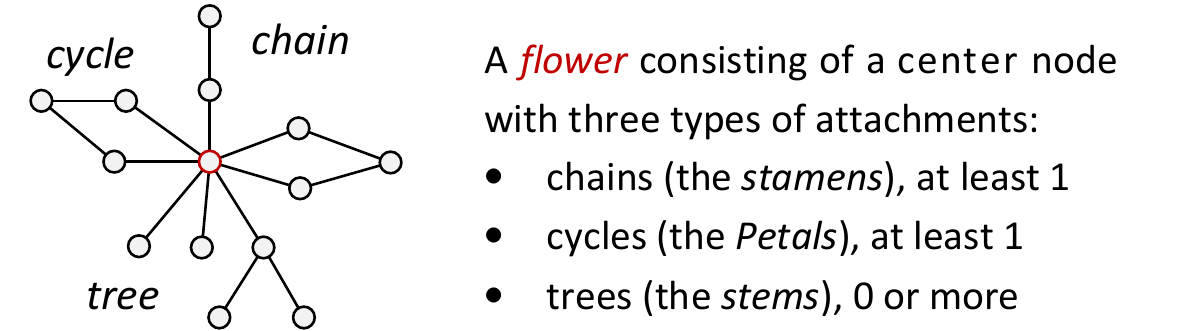}
    \caption{An example of \textit{flower} shape query graph}
    \label{fig:shape}
  \end{minipage}
\end{figure*}

\vspace{-0.12cm}
\subsection{Robustness with respect to Noise}
\label{scalability}
\vspace{-0.02cm}
Since we assume that different users may construct different query graphs to represent the similar query intention, we investigate the impact of varying query graphs on the performance of SGQ. To construct the semantically similar but structurally different graphs, we systematically considered node noise and edge noise. (1) \textit{\textbf{Node noise}}. We added the node noise by changing the node name or type with a randomly selected synonym or abbreviation. (2) \textit{\textbf{Edge noise}}. We added the edge noise by replacing the predicate with one of its top-10 semantically similar predicates in the predicate semantic space $\bm{E}$. (3) \textit{\textbf{Noise percentage}} is defined as the fraction of query graphs selected to add noise (varied from 10\% to 40\%). Figure \ref{fig:noise} shows the results for DBpedia (top-$k$=100): (1) All effectiveness metrics decrease as the noise ratio increases, (2) SGQ is more sensitive to edge noise. This is because SGQ may misunderstand the query intention if an inappropriate predicate is given. For example, if we use
\textit{designer} to replace \textit{assemble} in query Q117, then \textit{Automobiles} designed by Germans would be superior to \textit{Automobiles} assembled in \textit{Germany}. Furthermore, Table \ref{tab:noise} shows that the response time increases slightly with the growth of noise and it is sensitive to edge noise too.



\subsection{Scalability}
\label{scale}
This experiment studies the scalability of SGQ. We extracted two subgraphs $G_1$ and $G_2$ from DBpedia, e.g., $G_1$ has 3M nodes and 13.6M edges. Table \ref{tab:scale} shows the response time of SGQ for top-$k$=$\{$80,100,120$\}$ and the knowledge graph embedding time and memory usage. Observe that the time of SGQ increases as the graph size increases, but the change is not significant, which means that SGQ is scalable to the data size. This is because our approach can prune the unpromising candidates effectively for the different scale of the dataset. Moreover, our offline knowledge graph embedding time and memory usage are modest, e.g., within 6.6 hours and 8.8 GB.

\begin{table}
\setlength{\abovecaptionskip}{0.05cm}
\setlength{\belowcaptionskip}{-0.5cm}
\setstretch{0.8}
\fontsize{6.8pt}{2.8mm}\selectfont
  \centering
  \caption{Scalability evaluation over DBpedia}
  \begin{tabular} {p{1.9cm}<{\centering}||c|c|c||c|c}
   \hline
   \multirow{2}*{\textbf{(\#Nodes, \#Edges)}} & \multicolumn{3}{c||}{SGQ: \textit{online} (ms)} & \multicolumn{2}{c}{KG embedding: \textit{offline}}\\
   \cline{2-6}
    & $k$=$80$ & $k$=$100$& $k$=$120$ & time (h) & mem (GB)\\
   \hline \hline
   $G_1$(2M,9.8M) & 71.8 & 85.3 & 118.4 & 2.9 & 3.2\\
   \hline
   $G_2$(3M,13.6M) & 73.3 & 91.2 & 121.9 & 4.7 & 4.6\\
   \hline
   $G$(4.5M,15M) & 81.4 & 102.2 & 136.8 & 6.6 & 8.8\\
   \hline
  \end{tabular}
  \vspace{1mm}
  \label{tab:scale}
\end{table}

\begin{table}
\setlength{\abovecaptionskip}{0.05cm}
\setlength{\belowcaptionskip}{0.3cm}
\setstretch{0.8}
\fontsize{6.8pt}{2.8mm}\selectfont
\centering
  \caption{Effect of $\hat{n}$ and $\tau$ (DBpedia, top-$k$=100)}
\begin{tabular}{c||c|c|c|c||c|c|c|c}
    \hline
    \multirow{2}*{\textbf{Metrics}} &
    \multicolumn{4}{c||}{desired path length $\bm{\hat{n}}$} & \multicolumn{4}{c}{$pss$ threshold $\bm{\tau}$}\\
    \cline{2-9}
    & 2 & 3 & \textbf{4} & 5 & 0.6 & 0.7 & \textbf{0.8} & 0.9 \\ \hline
    Precision & 0.84 & 0.93 & \textbf{0.96} & 0.96 & 0.96 & 0.96 & \textbf{0.96} & 0.68 \\ \hline
    Recall & 0.42 & 0.45 & \textbf{0.48} & 0.48 & 0.48 & 0.48 & \textbf{0.48} & 0.34 \\ \hline
    F1 & 0.52 & 0.58 & \textbf{0.59} & 0.59 & 0.59 & 0.59 & \textbf{0.59} & 0.46 \\ \hline
    Time (ms) & 95.4 & 97.5 & \textbf{102.2} & 143.5 & 136.1 & 111.6 & \textbf{102.2} & 99.5 \\ \hline
\end{tabular}
\label{tab:pathlength}
\end{table}

\subsection{Parameter Sensitivity}
\label{parameter}
In this experiment, we evaluated the effect of user desired path length $\hat{n}$ and the path semantic similarity ($pss$) threshold $\tau$ on SGQ. First, we fixed $\tau$=0.8 and varied $\hat{n}$ from 2 to 5. Table \ref{tab:pathlength} shows that the effectiveness results are the same for $\hat{n}$=$\{4,5\}$, because all correct answers are defined as the schemas within 4-hop paths. As $\hat{n}$ decreases, less correct answers are found. Moreover, a larger $\hat{n}$ indicates that we need consider more search space, leading to longer response time. Second, we fixed $\hat{n}$=4 and varied $\tau$ from 0.6 to 0.9. The results show that the greater $\tau$ improves the response time, because more candidate paths are pruned. In addition, the effectiveness decreases when $\tau$=$0.9$, because the correct answers having the $pss$ between 0.8 and 0.9 are mistakenly pruned by a larger $\tau$.

\vspace{-0.1cm}
\section{Conclusions}
In this paper, we proposed a semantic-guided and response-time-bounded graph query to search knowledge graphs effectively and efficiently. We leveraged a knowledge graph embedding model to build the semantic graph for each query graph. Then we presented an A* semantic search to find the top-k semantically similar matches from the semantic graph according to the path semantic similarity. We optimized the A* semantic search to trade off the effectiveness and efficiency within a user-specific time bound, thereby improving the system response time. The experimental results on real datasets confirm the effectiveness and efficiency of our approach.

\vspace{-0.1cm}
\bibliographystyle{IEEEtran}
\bibliography{RDF}

\end{document}